\documentclass[useAMS,usenatbib]{mn2e}

\usepackage{txfonts}
\usepackage[dvips]{graphicx}
\usepackage{epsfig}
\usepackage{natbib}

\usepackage{multirow}
\usepackage{url}
\bibpunct{(}{)}{;}{a}{}{,}
\usepackage{color}
\usepackage{listings}

\title[The evolution of the galaxy content of dark matter haloes]%
{The evolution of the galaxy content of dark matter haloes}

\author[S.~Contreras et al.]
  {
 S.~Contreras$^{1,2}$, I.~Zehavi$^{3,2}$, C.~M.~Baugh$^{2}$, N.~Padilla$^{1,4}$, P.~Norberg$^{2,5}$
\\
 $^{1}$Instit\'uto Astrof\'{\i}sica, Pontifica
 Universidad Cat\'olica de Chile, Santiago, Chile \\
 $^{2}$Institute for Computational Cosmology, Department of Physics, 
Durham University, South Road, Durham, DH1 3LE, UK \\
 $^{3}$Department of Astronomy, Case Western Reserve University, Cleveland, OH
44106, USA \\  
 $^{4}$Centro de Astro-Ingenier\'{\i}a, Pontificia 
 Universidad Cat\'olica de Chile, Santiago, Chile \\
 $^{5}$Centre for Extragalactic Astronomy, Department of Physics, 
Durham University, South Road, Durham, DH1 3LE, UK \\
}

\def\LaTeX{L\kern-.36em\raise.3ex\hbox{a}\kern-.15em
    T\kern-.1667em\lower.7ex\hbox{E}\kern-.125emX}

\begin{document}

\pagerange{\pageref{firstpage}--\pageref{lastpage}} \pubyear{2014}

\maketitle

\label{firstpage}

\begin{abstract}
We use the halo occupation distribution (HOD) framework to characterise 
the predictions from two independent galaxy formation models  
for the galactic content of dark matter haloes and its evolution 
with redshift. Our galaxy samples correspond to a range of fixed number 
densities defined by stellar mass and span $0 \le z \le 3$.
We find remarkable similarities between the model predictions. Differences 
arise at low galaxy number densities which are sensitive to the treatment 
of heating of the hot halo by active galactic nuclei. 
The evolution of the form of the HOD can be described in a relatively 
simple way, and we model each HOD parameter using its value at $z=0$
and an additional evolutionary parameter.  In particular, we find that the ratio
between the characteristic halo masses for hosting central and satellite 
galaxies can serve as a sensitive diagnostic for galaxy evolution models. 
Our results can be used to test and develop empirical studies of galaxy evolution and can facilitate the
construction of mock galaxy catalogues
for future surveys.
\end{abstract} 


\section{Introduction}
\label{Sec:Intro}

In the standard cosmological framework, galaxies form, evolve and reside
in dark matter haloes. A key requirement of this framework is to understand
how galaxies populate dark matter haloes. What determines how many galaxies 
are hosted by a dark matter halo?  How do the properties of galaxies depend
on the mass of the halo?  These questions lie at the core of galaxy formation
theory. The answers are also crucial if we are to take full advantage of 
the next generation of galaxy surveys, which aim to make pristine clustering 
measurements to pin down the nature of dark energy. The extraction of 
cosmological inferences from these data will no longer be dominated by 
statistical  errors but instead will be limited by the accuracy of our 
theoretical models. Understanding how galaxies relate to the underlying 
dark matter is thus essential for optimally utilizing the large-scale 
distribution of galaxies as a cosmological probe.

The clustering of dark matter is dominated by gravity and can be computed
reliably with cosmological N-body simulations. However, the detailed physics 
of galaxy formation -- gas cooling, star formation, and feedback effects -- is 
only partially understood, so that the relation between galaxies and the
underlying dark matter cannot be predicted robustly from first principles.

A useful approach to study this is semi-analytic modeling (SAM) of galaxy 
formation (for reviews see, e.g., \citealt{Cole:2000,Baugh:2006,Benson:2010,Somerville:2015}). In such models, haloes identified in N-body simulations are 
``populated'' with galaxies using analytical prescriptions for the baryonic 
processes.  Following the dark matter merger trees, galaxies merge and evolve 
as new stars form and previous generations of stars change. 
Different feedback or heating mechanisms, such as those caused by star 
formation, active galactic nuclei, or the photo-ionizing ultra-violet 
background, are also incorporated.  SAMs have been successful in reproducing 
a range of observed properties including stellar mass functions and galaxy 
luminosity functions 
(see, e.g., 
\citealt{Bower:2006,Croton:2006,Fontanot:2009,Guo:2011,Guo:2013a,Gonzalez:2014,Padilla:2014,Henriques:2015,Lacey:2016,Croton:2016}).

The connection between the mass of a dark matter halo and the galaxies 
which populate it is often expressed through the halo occupation 
distribution (HOD) framework (e.g., 
\citealt{Jing:1998a,Benson:2000,Peacock:2000,Seljak:2000,Scoccimarro:2001,Berlind:2002,Berlind:2003,Cooray:2002,Yang:2003,Kravtsov:2004,Zheng:2005,Conroy:2006}). The HOD formalism describes the ``bias''
relation between galaxies and mass at the level of individual dark matter
haloes, in terms of the probability distribution that a halo of virial mass
$M_{\rm h}$ contains $N$ galaxies which satisfy a particular selection criterion.  
It transforms 
measures of galaxy clustering into a physical relation between galaxies
and dark matter haloes, setting the stage for detailed tests of galaxy
formation models.
The HOD approach has proven to be a very powerful theoretical tool
to constrain the galaxy-halo connection and has been applied
to interpret clustering data from numerous surveys at low and high 
redshifts (e.g., \citealt{Jing:1998b,Jing:2002,Bullock:2002,Moustakas:2002,vandenBosch:2003,Magliocchetti:2003,Yan:2003,Zheng:2004,Yang:2005,Zehavi:2005,Zehavi:2011,Cooray:2006,Hamana:2006,Lee:2006,Lee:2009,Phleps:2006,White:2007,White:2011,Zheng:2007,Zheng:2009,Blake:2008,Brown:2008,Quadri:2008,Wake:2008,Wake:2011,Kim:2009,Simon:2009,Ross:2010,Coupon:2012,Coupon:2015,delaTorre:2013,Krause:2013,Parejko:2013,Guo:2014b,Durkalec:2015,Kim:2015,McCracken:2015,Skibba:2015}).

HOD models have mostly been used to constrain the relation between 
galaxies and haloes at a fixed epoch, as the HOD approach by itself does not 
offer any guidance as to how to treat the evolution of the galaxy population 
over cosmic time. Attempts to study galaxy evolution using this framework have 
for the most part explored ``snapshots'' of 
clustering at different epochs to empirically constrain the evolution (e.g., \citealt{Zheng:2007,White:2007,Wake:2008,Wake:2011,Abbas:2010,Coupon:2012,delaTorre:2013,Guo:2014b,Manera:2015,Skibba:2015}).
but a complete model for the evolution of the HOD is still missing. 

Our goal is to remedy this situation and develop a theoretical
model for this evolution by studying how the HOD changes with time. 
A simplified approach in this vein was taken by \citet{Seo:2008} who studied
the predictions for passive evolution of the HOD by populating simulations with galaxies according to a range of assumed HOD and tracking their evolution
with time. That work is of limited use due to the unphysical assumption of 
passive evolution. The form of the HOD at different redshifts has also been 
studied in the context of abundance matching modeling
\cite{Kravtsov:2004,Conroy:2006}.
Here we will perform a comprehensive study of the
evolution of the HOD using the predictions of semi-analytic modeling which 
captures the important galaxy formation physics.  

The present paper builds upon our work exploring the predictions of 
SAM of galaxy formation, focusing on the connection 
between galaxies and their host dark matter haloes. \cite{C13} demonstrated 
that SAMs from different groups give consistent predictions for the galaxy 
correlation function on large scales, for samples constructed to have the same 
abundance of galaxies, and that the differences on small scales (the so-called 
one-halo term) can be readily understood in terms of the choices made about the placement of galaxies 
within dark matter haloes. In a second paper, we examined the connection between 
different galaxy properties and the mass of the dark matter halo hosting the 
galaxy \citep{C15}. We found that some properties, such as stellar mass, 
depend on subhalo mass in a monotonic fashion (albeit with a scatter), whereas others, such as the cold gas mass, have a more complex dependence on halo mass. 

Here we use the HOD formalism to compare how different models populate 
haloes with galaxies over cosmic time. 
We study the output of two independently developed SAMs, originally from the 
Durham and Munich groups, at different number densities. This allows us to 
assess which features of the 
predicted HODs are generic and which are sensitive to the details of the 
modelling of the various physical processes, and how best 
to describe the evolution of the HOD at a given number density. 
We also consider some simplified empirical models for the evolution of 
the galaxy distribution and show how these differ from the predictions of the 
SAMs. 

This study will enable the incorporation of evolution into the halo models, an 
aspect that is absent from the standard implementation. The applications of 
this are two fold. First, from the observational side, it will allow for
a consistent combined analysis of clustering measures over a range of epochs,
in order to constrain galaxy formation and evolution. A second application 
of the HOD is to quantify evolution in the galaxy population, which will
facilitate the creation of realistic 
mock galaxy catalogues for surveys which span a large range of lookback times. 
Accurate estimates exist for the form of 
the HOD using measurements of the galaxy clustering in, for example, local 
surveys (e.g. \citealt{Zehavi:2011}). These can be used, in conjunction with 
a sample of dark matter haloes extracted from an N-body simulation to build 
a mock catalogue with the same clustering properties and abundance of galaxies.
The problem then is how to extend this approach to build a catalogue that 
expands beyond the redshift interval covered by the original survey, for use 
with upcoming surveys. 
Our evolution study presented here is an essential input for such efforts.

The outline of this paper is as follows: in Section~\ref{Sec:GC} we introduce 
the SAMs used, along with the N-body simulation the models are grafted onto, 
and we describe the HOD characterisation of the galaxy population. In 
Section~\ref{Sec:HOD_Ev} we show the evolution of the HODs for a wide range 
of number densities and redshifts, we fit the HODs predicted by the SAMs and 
we show the evolution of the best-fitting parameters. In Section~\ref{Sec:PC2} 
we compare our results with simple models for the evolution of galaxy 
clustering. Finally in Section~\ref{Sec:Concl} we present our conclusions.

\section{Models of Galaxy Clustering} 
\label{Sec:GC}
Here we review the different galaxy formation 
models used (Section~\ref{SubSec:GFM}) and outline the halo occupation 
distribution (HOD) description of the way in which dark matter haloes are 
populated by galaxies (Section~\ref{SubSec:HOD}).

\subsection{Galaxy formation models}
\label{SubSec:GFM}
We first give a brief overview of the galaxy formation models we use (\S~\ref{SubSubSec:SAM}). We then provide the details of the N-body simulation they are 
implemented in and outline the construction of the halo merger 
trees and reconcile the different halo mass definition used by the groups of modellers 
(\S~\ref{SubSubSec:Mill}).

\subsubsection{The semi-analytic models}
\label{SubSubSec:SAM}

The objective of SAMs is to model the main physical processes involved in the evolution and formation of galaxies. Some of these processes are:
collapse and merging of dark matter haloes; shock heating and radiative 
cooling of gas; star formation; supernovae, AGN, and photoionisation feedback;
chemical enrichment of gas and stars; disc instability; and galaxy mergers.

We chose SAMs that have different implementations of these processes, 
so that we can identify which results are robust and which ones depend on 
the physical treatment in the models.

The SAMs we use are those of \cite{Guo:2013a} (hereafter G13) and 
\cite{Gonzalez:2014} (hereafter GP14)\footnote{The G13 and GP14 outputs 
are publicly available from the Millennium Archive in Durham \url{http://virgodb.dur.ac.uk/}
and Garching \url{http://gavo.mpa-garching.mpg.de/Millennium/}}. 
The G13 model is a  version of {\tt L-GALAXIES}, the SAM code of the 
Munich group  \citep{DeLucia:2004,Croton:2006,DeLucia:2007,Guo:2011,Henriques:2013,Henriques:2015}. 
The GP14 model is a version of {\tt GALFORM}, developed by the 
Durham group \citep{Bower:2006,Font:2008,Lagos:2012,Lacey:2016}. 
The GP14 model has an improved treatment of star formation, dividing  
the interstellar medium into molecular and atomic hydrogen components 
(which was introduced by \citealt{Lagos:2011b}). An important 
improvement of G13 and GP14 over their immediate predecessors 
(\citealt{Guo:2011,Lagos:2012} respectively), is the use of a recent   
cosmological simulation with an updated cosmology (specified below). 
One notable difference between the G13 and GP14 models is the 
treatment of satellite galaxies. In GP14, a galaxy is assumed to lose 
all its hot gas halo and start decaying onto the central galaxy once 
it becomes a satellite.  In G13, these processes are more gradual and 
depend on the orbit of the satellite and the destruction of the subhalo 
\citep{Font:2008}.

\subsubsection{The Millennium simulations and halo merger trees}
\label{SubSubSec:Mill}

The SAMs we consider are implemented in the same N-body simulation, 
the Millennium-WMAP7 run (hereafter MS7; G13, GP14) which is similar 
to the original 
Millennium simulation \citep{Springel:2005} but uses a WMAP7 
cosmology\footnote{The values of the cosmological parameters used in 
the MS7 are: $\Omega_{m0}$= $\Omega_{dm0}$+$\Omega_{b0}$ = 0.272, $\Omega_{\Lambda0}$ = 0.728, $\Omega_{b0}$ = 0.0455, $\sigma_8$ = 0.81, $n_{\rm s}$ = 0.967, $h$ = 0.704.}. 
The simulation uses $2160^3$ particles in a $(500\, h^{-1}\, {\rm~Mpc})^3$ box,
corresponding to 
a particle mass of $9.31\times10^8 h^{-1} \,{\rm M_{\odot}}$. There are 61 simulation 
outputs between $z = 50$ and $z=0$. 

Halo merger trees are constructed from the simulation outputs,
These trees are the starting point for the SAMs. Both the G13 and GP14 use a 
friends-of-friends ({\tt FoF}) group finding 
algorithm \citep{Davis:1985} to identify haloes in each snapshot of 
the simulation, retaining those with at least 20 particles. 
{\tt SUBFIND} is then run on these groups to identify subhaloes 
\citep{Springel:2001}. 
The merger trees differ from this point on. 
G13 construct dark matter halo merger trees by linking a subhalo 
in one snapshot to a single descendant subhalo in the subsequent output. 
The halo merger tree used in {\tt L-GALAXIES} is therefore a subhalo merger
tree. GP14 use the {\tt Dhalo} merger tree construction
\citep{Merson:2013,Jiang:2014} that also uses 
the outputs of the {\tt FoF} and {\tt SUBFIND} algorithms. 
The {\tt Dhalo} algorithm applies conditions on the amount of mass 
stripped from a subhalo and its distance from the centre of the 
main halo before it is considered to be merged with the main subhalo. 
Also, subsequent output times are examined to see if the subhalo moves 
away from the main subhalo, to avoid merging subhaloes which have merely 
experienced a close encounter before moving apart. {\tt GALFORM} 
post-processes the {\tt Dhalo} trees to ensure that the halo mass increases 
monotonically with time.

The definition of halo mass used in the two models is not the same. The
{\tt Dhalo} mass used in {\tt GALFORM} corresponds to an integer number of
particle masses whereas a virial mass is calculated in {\tt L-GALAXIES}. This 
leads to differences in the halo mass function between the models.  
To compare the HODs predicted by the models we need a common 
definition of halo mass. \cite{Jiang:2014} showed that the {\tt Dhalo} 
masses and virial halo masses can be related by applying a small offset in mass 
and a scatter. For simplicity, we chose instead to relabel the halo masses 
in GP14 by matching the abundance of dark matter haloes between models 
and using the mass from G13.

\subsection{The halo occupation distribution}
\label{SubSec:HOD}

\subsubsection{HOD modelling}
\label{SubSubSec:HOD}

The HOD formalism  characterises the relationship between galaxies 
and haloes in terms of the probability distribution that a halo 
of virial mass $M_{\rm h}$ contains $N$ galaxies of a given type, together with 
the spatial and velocity distributions of galaxies inside haloes. The key
ingredient is the halo occupation function, $\langle N(M_{\rm h}) \rangle$, which 
represents the average number of galaxies as a function of halo mass. 
The advantage of this approach is that it does not rely on assumptions 
about the (poorly understood) physical processes that drive galaxy formation
and can be empirically derived from the observations.  

Standard applications typically assume a cosmology as well as a parametrized 
form for the halo occupation functions motivated by the predictions of 
SAMs and hydrodynamics simulations (e.g., \citealt{Zheng:2005}). The HOD 
parameters are then constrained using measurements of galaxy clustering 
measurements from large surveys and the theoretically predicted halo 
clustering. This approach essentially transforms 
measures of galaxy clustering into a physical relation between galaxies and 
dark matter haloes, setting the stage for detailed tests of galaxy formation 
models.  

An important application of HOD modelling is to facilitate the generation of 
mock galaxy catalogues by populating dark matter haloes from an N-body 
simulation with galaxies that reproduce the desired clustering properties. 
This method has become popular due to its low computational cost
and good performance (e.g., \citealt{Manera:2015,Zheng:2016}, Smith et al., 
in prep.). Typically, the halo occupation function 
is available for an observational sample at a particular redshift, 
or over a narrow redshift interval. In order to generate a mock galaxy 
catalogue over a wide baseline in redshift using this technique, it is 
necessary to specify the HOD as a function of redshift or to have a 
prescription for its redshift evolution. 
SAMs predict the galaxy content of haloes and so the HOD is an output of 
these models. Here we use the HOD to describe the model 
predictions at different redshifts for galaxy samples with different 
abundances. 

\begin{figure}
\includegraphics[width=0.48\textwidth]{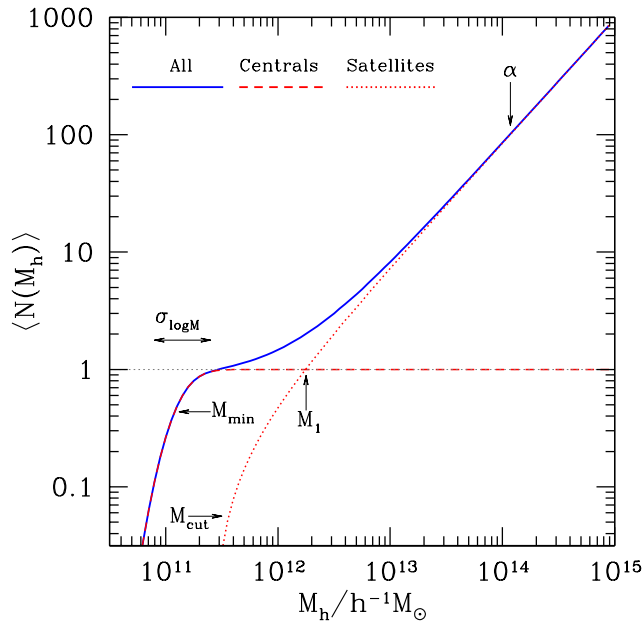}
\caption{
A schematic depicting the standard 5-parameter form of the HOD, 
which gives the mean number of galaxies per halo as a function of host 
halo mass.  The solid blue line shows the occupation function for all 
galaxies, which can be broken up into the contribution from central 
galaxies (red dashed line) and satellite galaxies (red dotted line). 
To guide the eye a grey dotted line is plotted at 
$\langle N(M_{\rm h})\rangle = 1$; 
this will be shown in all subsequent HOD plots. 
The halo occupation function of central galaxies shows a gradual transition 
from zero to one galaxy per halo and can be well described by two parameters; 
$\sigma_{\log M}$, which describes the smoothness of this transition and 
$M_{\rm min}$, which is the halo mass at which half of the haloes 
are populated by a central galaxy. The satellites occupation function is 
described as a transition from zero galaxies to a power law and is 
characterised by three parameters: $M_{\rm cut}$ the minimum halo mass 
at which satellites first populate dark matter haloes; $M_{1}$ the mass where there 
is an average of one satellite galaxy per halo; and $\alpha$,\ the power 
law slope. For a full description of these parameters see 
Section~\ref{SubSubSec:Param}.}
\label{Fig:Guia}
\end{figure}

\subsubsection{HOD parametrization}
\label{SubSubSec:Param}

For the parametrization of the HOD, it is useful to make a distinction
between central galaxies, namely the main galaxy inside a halo, and
the additional satellite galaxies that populate the halo, and consider
separately the contributions of each
\citep{Kravtsov:2004,Zheng:2005}.
By definition, a dark matter halo cannot be populated by more than one
central galaxy but, in principle, there is no limit to the number of 
satellite galaxies.  Also, for samples defined by properties that scale 
with the halo mass, such as luminosity, a halo should be first populated 
by a central galaxy and then by a satellite galaxy. (One counterexample to 
this is when a selection involving colour is applied, as is the 
case with luminous red galaxies.)
In SAMs, there is no unique way to define which is the central galaxy. 
Following a halo merger, {\tt L-GALAXIES} defines a central galaxy as the
most massive galaxy inside a halo in terms of the stellar mass. 
In {\tt GALFORM} the central galaxy is assumed to be the one from the most 
massive progenitor halo. Despite this distinction, both models usually agree 
in their identification of the central.

The traditional shape assumed for the HOD is a rapid transition from zero to
one galaxy for centrals and a transition from zero galaxies to a power law 
for satellite galaxies. One of the most commonly used parametrizations is  
the 5-parameter model introduced by \cite{Zheng:2005} (see also 
\citealt{Zheng:2007,Zehavi:2011}), which describes well samples of galaxies 
brighter than a given luminosity or more massive than a given stellar mass.  
Here we will adopt this form of the halo occupation function to describe the 
predictions of the SAMS.

The mean occupation function of the central galaxies is a step-like function
with a cutoff profile softened to account for the scatter between galaxy
luminosity and halo mass. It has the following form:

\begin{equation}
 \langle N_{\rm cen}(M_{\rm h})\rangle = \frac{1}{2}\left[ 1 + {\rm erf} \left( \frac{\log M_{\rm h} - \log M_{\rm min}}{\sigma_{\log M}}  \right) \right],
\label{Eq:Cen_HOD}
\end{equation}
where $M_{\rm h}$ is the host halo mass and 
$ {\rm erf}(x)$ is the error function,
\begin{equation}
 {\rm erf}(x) = \frac{2}{\sqrt{\pi}} \int_{0}^{x} e^{-t^2} {\rm d}t.
\end{equation}
$M_{\rm min}$ characterises the minimum halo mass for hosting a central galaxy
above the specified threshold. Its exact definition can vary between different
HOD parametrizations. In the form we adopt here, it is the halo mass  
for which half of the haloes host a central galaxy above a given threshold 
(ie. $\langle N_{\rm cen}(M_{\rm min})\rangle = 0.5$).  
The other parameter, $\sigma_{\log M}$, characterises the width  of the 
transition from zero to one galaxy per halo. A value of
$\sigma_{\log M}=0$  corresponds to a vertical step-function transition, 
while a non-zero value of $\sigma_{\log M}$ is indicative of the amount of 
scatter between stellar mass and halo mass.
For samples defined by a luminosity threshold, it was further shown that 
$M_{\rm min}$ is the mass of haloes in which the mean luminosity of central 
galaxies is the luminosity threshold, and $\sigma_{\log M}$ is directly 
related to the width of the distribution of central galaxy luminosities
\citep{Zheng:2005,Zheng:2007}.

For satellite galaxies, the HOD is modelled as:
\begin{equation}
 \langle N_{\rm sat}(M_{\rm h})\rangle = \left( \frac{M_{\rm h}-M_{\rm cut}}{M^*_1}\right)^\alpha,
\label{Eq:Sat_HOD}
\end{equation}
for $M_{\rm h}>M_{\rm cut}$,
representing a power-law occupation function modified by a smooth cutoff at
small halo masses.  Here $\alpha$ is the slope of the power-law, which 
typically has a value close to unity, 
$M_{\rm cut}$ is the satellite cutoff mass scale (i.e., the minimum mass of haloes
hosting satellites), and $M^*_1$ is the normalization of the power law.
A useful parameter that is often discussed is $M_{1}$, the mass of haloes
that on average have one satellite galaxy, defined by 
$\langle N_{\rm sat}(M_{1})\rangle = 1$. Note that $M_1$ is different from
 $M^*_1$ above, but is obviously related to the values of $M^*_1$ and 
$M_{\rm cut}$  
($M_1 = M_{\rm cut} + M^*_1$). 

The occupation functions for centrals and satellites can be fitted 
independently with this definition, with the total number of galaxies given 
by their sum: 
\begin{equation}
 \langle N_{\rm gal}(M_{\rm h})\rangle =  \langle N_{\rm cen}(M_{\rm h})\rangle +  \langle N_{\rm sat}(M_{\rm h})\rangle.
\end{equation}
A schematic representation of the shape of the HOD illustrating which 
features are 
sensitive to the various parameters is shown in Fig.~\ref{Fig:Guia}.

Other works have also applied the cutoff profile used for the central galaxies
occupation function to the satellites, effectively assuming (using our
notation) that the total number of galaxies is given by 
$\langle N_{\rm cen}\rangle(1+\langle N_{\rm sat}\rangle)$ 
(e.g., \citealt{Zheng:2007,Zheng:2009,Zehavi:2011,Guo:2015}). In this case 
the fitting of the HOD cannot be done separately for centrals and 
satellites (because of the $\langle N_{\rm cen} \rangle \langle N_{\rm sat} 
\rangle$ term). Hence, assuming this form results in a more complex 
procedure to determine the best-fitting  values of the parameters and 
ultimately gives poorer constraints, particularly for $M_{\rm cut}$. 
This assumption is often used when the HOD is inferred from the measured
projected correlation function. Caution must be taken before comparing results 
obtained with this formalism and the one presented here.

\begin{figure}
\includegraphics[width=0.48\textwidth]{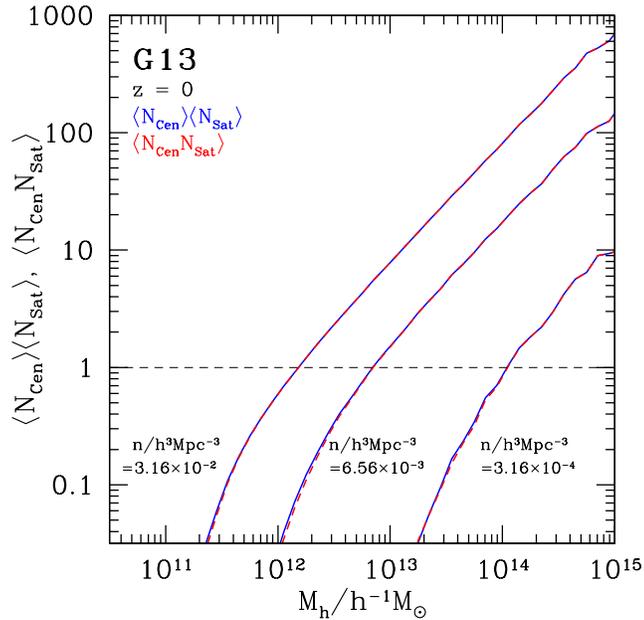}
\caption{A test of the assumption, often made in HOD modelling, that 
$\langle N_{\rm sat}\rangle \langle N_{\rm cen}\rangle$ (blue lines) 
and $\langle N_{\rm sat} N_{\rm cen}\rangle$ (red lines) are equivalent. 
These quantities are shown as a function of halo mass in the G13 model 
for three number densities: $n =  3.16\times 10^{-2}\, h^{3} \,{\rm Mpc}^{-3}$, 
$n =  6.56\times 10^{-3}\, h^{3}\, {\rm Mpc}^{-3}$ and 
$n =  3.16\times 10^{-4}\, h^{3}\, {\rm Mpc}^{-3}$ (moving from left to right 
in order of decreasing density)  at $z=0$.
These two quantities are indeed equivalent in the SAM output. 
}
\label{Fig:CenSat2}
\end{figure}

When inferring the HOD from measured galaxy clustering, one needs to specify 
the mean value of $\langle N_{\rm cen} N_{\rm sat} \rangle$ at each halo mass when 
computing the one-halo central--satellite pairs. It is often implicitly
assumed that a halo hosting a satellite galaxy also hosts a central galaxy 
from the same sample.  
Alternatively, one can assume independent central and satellite occupations and 
approximate this term as 
$\langle N_{\rm cen} \rangle \langle N_{\rm sat} \rangle$  (see discussion in 
\citealt{Zheng:2005,Guo:2014b,Guo:2015a}). 
The exact level of the correlation between central
and satellite galaxies is determined by galaxy formation physics.
While not relevant to our direct computations of the HOD in the SAMs in this 
paper,   we can use the output of the SAMs to assess the impact of the
central-satellite correlation and test the assumption that 
$\langle N_{\rm cen} N_{\rm sat}\rangle \simeq \langle N_{\rm cen}\rangle 
\langle N_{\rm sat}\rangle$. 
Fig.~\ref{Fig:CenSat2} shows these two quantities plotted as a function 
of halo mass for samples with different galaxy number densities using the output of the G13 SAM.
We find only negligible difference between these two terms for small 
occupation values, demonstrating that this 
approximation is indeed valid for these number densities.

\begin{figure}
\includegraphics[width=0.48\textwidth]{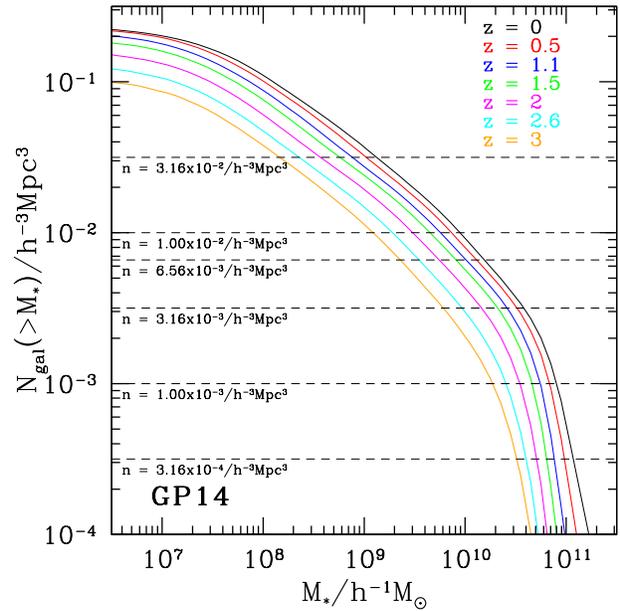}
\includegraphics[width=0.48\textwidth]{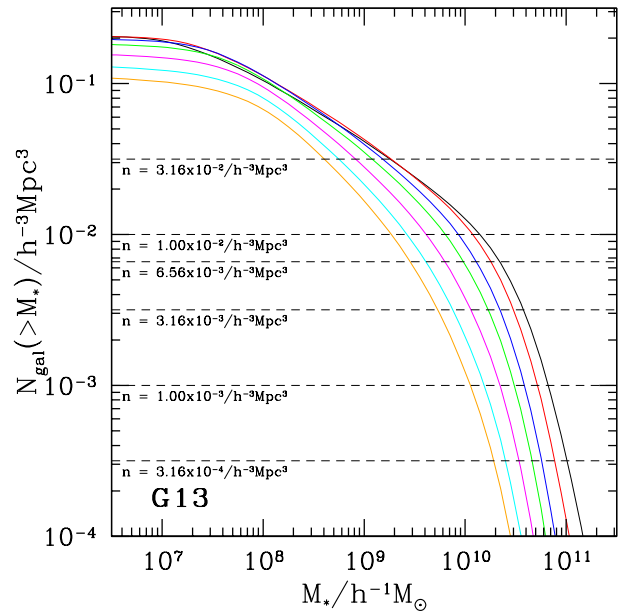}
\caption{
The cumulative stellar mass function in the GP14 (top) and G13 (bottom) models. 
The different lines represent different redshifts as labelled in the 
top panel, with the redshift increasing from top to bottom.
The dashed horizontal lines show the number density cuts adopted, which are labelled 
by the value of the number density. The galaxies selected for a given number density are those 
to the right of the intersection with their associated dashed line. }
\label{Fig:ASMF}
\end{figure}

\section{The evolution of the HOD}
\label{Sec:HOD_Ev}

This section contains our main results. In Section~\ref{SubSec:ND} we 
introduce the galaxy samples studied and plot the HODs predicted for these 
samples by the SAMs. In Section~\ref{SubSec:zEv} we show how the HODs evolve 
with redshift. Fits to the HODs are given in Section~\ref{SubSec:Fit}. 
In Section~\ref{SubSec:FitEv} we quantify the evolution of the best-fitting 
parameters. Finally, in Section~\ref{SubSec:M1Min} we show the evolution of 
the ratio between $M_1$ and $M_{\rm min}$, two of the HOD parameters often used 
to characterise the galaxy-halo relation.

\subsection{HOD for galaxy samples with different number densities}
\label{SubSec:ND}

For the main part of our work we use the number density of galaxies
ranked in order of decreasing stellar mass to define our galaxy samples
in the SAM catalogues. We build galaxy samples for a wide range of number 
densities: $n = 3.16\times 10^{-2}, 1\times 10^{-2}, 6.56\times 10^{-3}, 
3.16\times 10^{-3}, 1\times 10^{-3}$ and $ 3.16\times 10^{-4} \,h^{3}\, 
{\rm Mpc}^{-3}$ and redshifts: $z=0, 0.5, 1.1, 1.5, 2, 2.6$ and $3$.
The cumulative comoving number density of galaxies as a function of stellar
mass is commonly used to try to link galaxy populations across cosmic time
(\citealt{Leja:2013,Mundy:2015,Torrey:2015}; see also our discussion in 
\S~\ref{Sec:PC2}). 
It is inspired by the same hypothesis that motivates the
passive evolution model (\S~\ref{Sec:PC2}), is better defined 
than a constant stellar mass selected sample, being insensitive 
to systematic shifts in the stellar mass calculation, and is 
readily reproducible in observations. 
\cite{C13} have also shown that HOD predictions for such samples are quite 
robust among different SAMs at a fixed redshift.
While the samples in this work are all ranked by stellar mass,
our main results regarding the HOD evolution also hold when 
defining galaxy samples using other galaxy properties that scale with the halo 
mass, e.g., luminosity. 

The samples were chosen to be evenly spaced in logarithmic number density, with 
each one corresponding to a change of half a decade in log 
abundance. There are three densities in particular which we use to illustrate 
our main results: 
(i)  $3.16\times 10^{-2} \,h^{3} \, {\rm Mpc}^{-3}$, which is close to the 
number density studied by \cite{Zheng:2005} in their comparison between 
SAMs and hydrodynamical simulations, 
(ii) $6.56\times 10^{-3} \, h^{3} \, {\rm Mpc}^{-3}$ which is roughly the number 
density of galaxies brighter than $L^*$  in the SDSS Main Sample 
\citep{Zehavi:2011}, and
(iii) $3.16\times 10^{-4} \, h^{3}\, {\rm Mpc}^{-3}$, which is comparable to the 
number density of luminous red galaxies in the SDSS-III Baryonic Oscillation 
Spectroscopic Survey \citep{Eisenstein:2011}.

Fig.~\ref{Fig:ASMF} shows the cumulative stellar mass function for all
redshifts studied. The horizontal dashed lines show the different number 
density cuts we consider. The galaxies selected for a given number
density are those to the right of the intersection with their
associated dashed line. For ease of reference, we specify the corresponding
stellar mass thresholds at $z=0$ later on in Tables 1 and 2.

\begin{figure}
\includegraphics[width=0.48\textwidth]{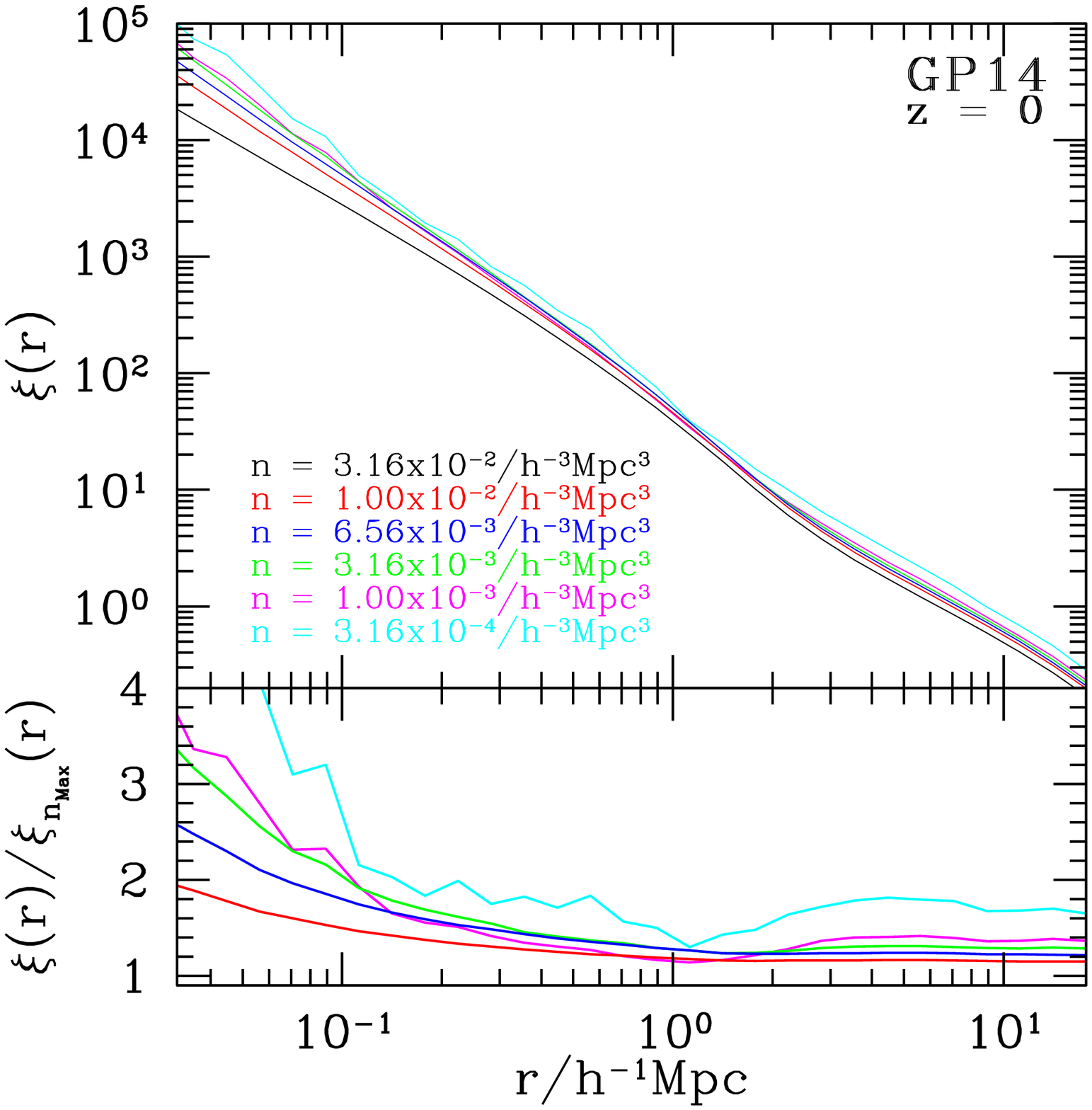}
\includegraphics[width=0.48\textwidth]{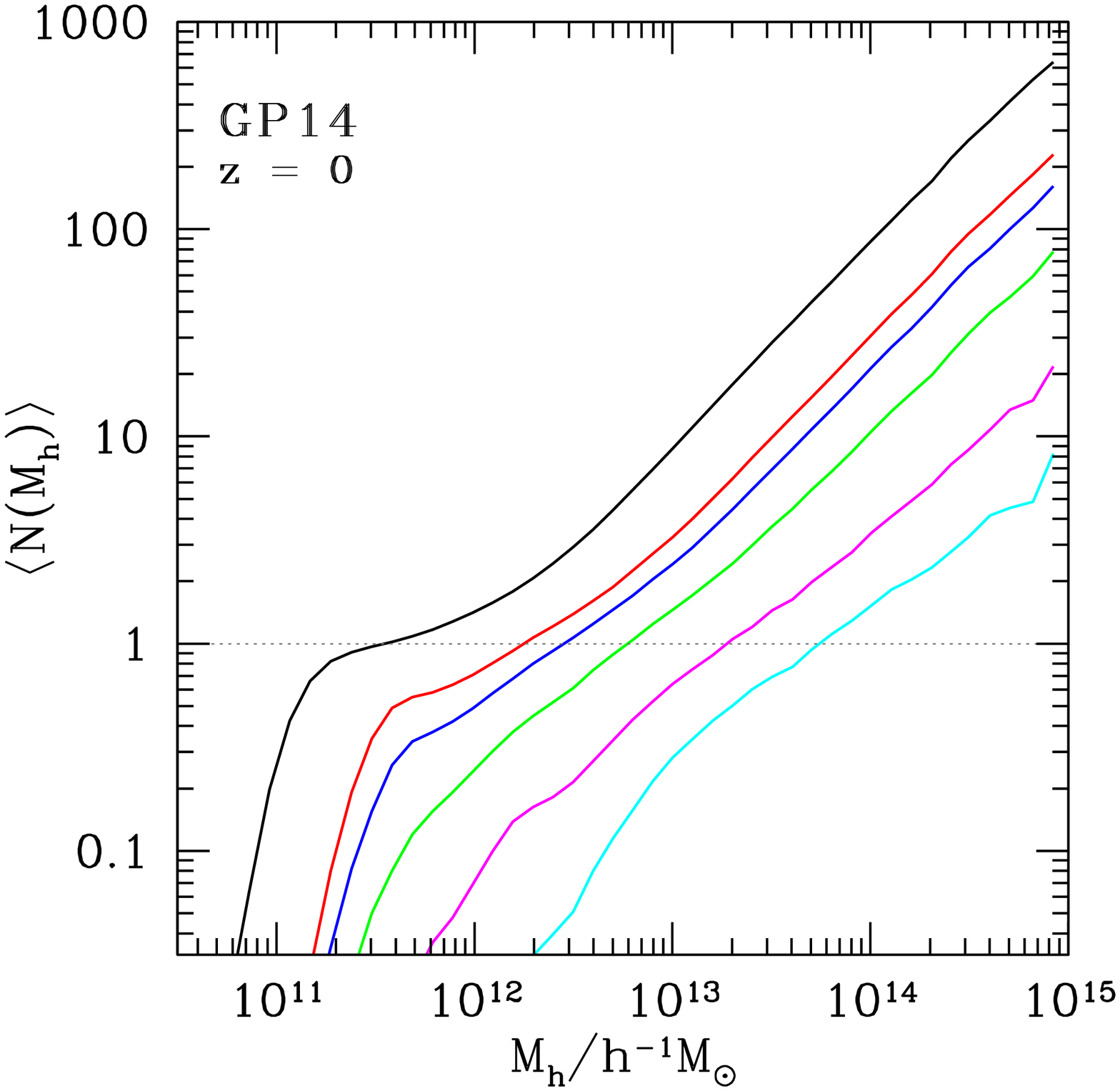}
\caption{The correlation function (top panel) and HOD (bottom panel)
at $z=0$ for the GP14 model. The different colors represent different
number densities as labelled on the top panel. 
In the bottom part of the top panel we show the ratio between the 
correlation functions for different number densities and the one corresponding
to the highest number density ($n = 3.16\times 10^{-2} h^{3} {\rm Mpc}^{-3}$).
In the top panel, the number density increases from top to bottom.
In the bottom panel, the number density decreases from top to bottom.
}
\label{Fig:n_ev}
\end{figure}

Fig.~\ref{Fig:n_ev} shows the correlation functions and halo occupation 
functions calculated directly in the GP14 SAM,  for all number densities 
at $z=0$. 
The clustering amplitude correlation increases monotonically as the galaxy
number density decreases (corresponding to more massive, or more luminous,
galaxies). 
The HOD shifts toward higher halo masses as the galaxy number density 
decreases, which means that more massive galaxies occupy more massive haloes.
Also, with decreasing number density, the transition between zero and one 
central galaxy per halo becomes broader which makes the plateau in the 
halo occupation where it is dominated by central galaxies less pronounced. 
Similar results are found for G13,  and the implications for the 
HOD parameters (for both) are shown and discussed in Section~\ref{SubSec:Fit}).
Identical trends were found by \cite{Zehavi:2011} using the SDSS for samples
of varying luminosity thresholds.

\subsection{The evolution of the HOD with redshift}
\label{SubSec:zEv}

\begin{figure*}
\includegraphics[width=0.96\textwidth]{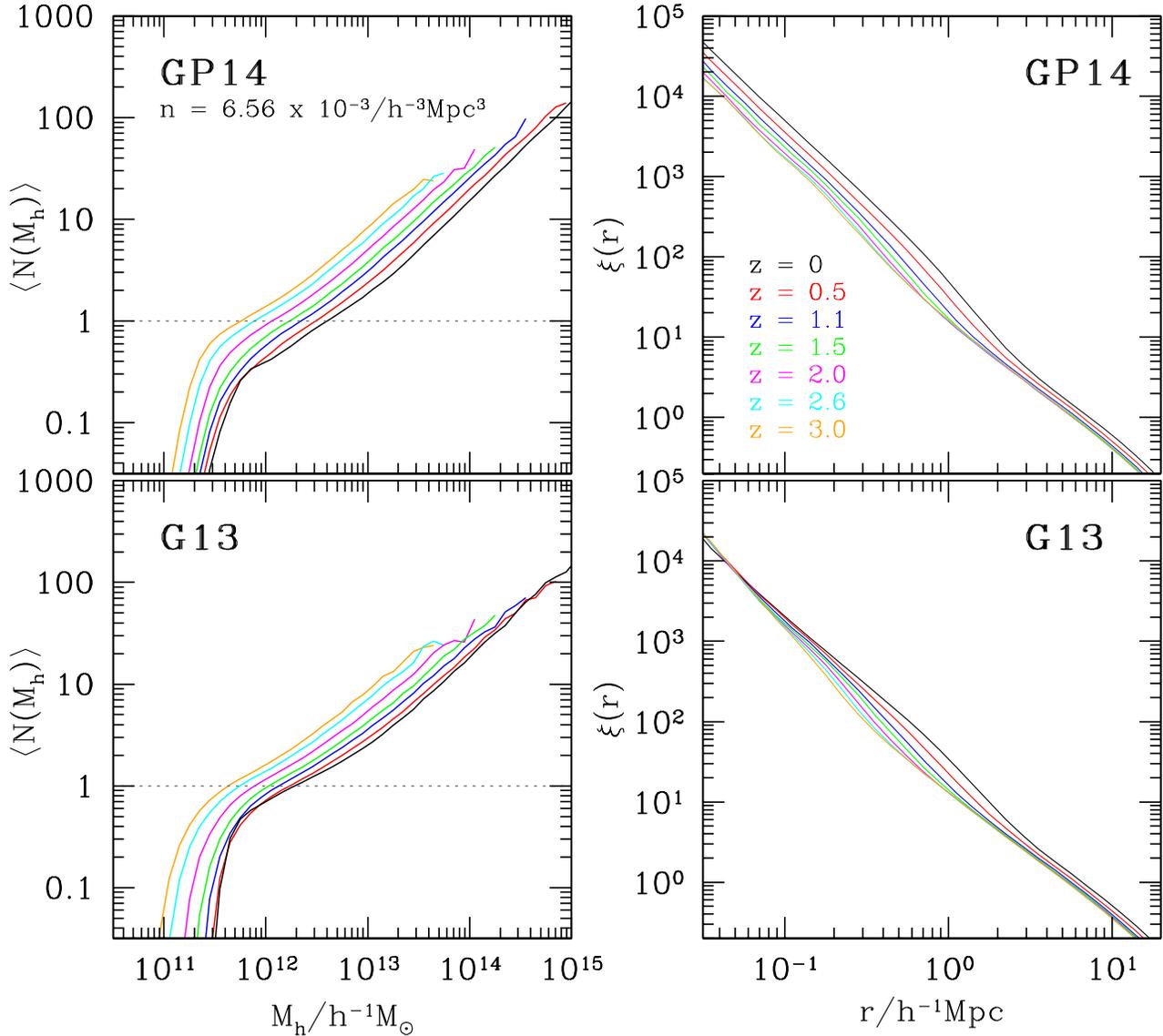}
\caption{
The evolution with redshift of the HOD (left panel) and correlation function (right panel) 
for a galaxy sample with number density $n = 6.56 \times 10^{-3} \, h^{3} \, {\rm Mpc}^{-3}$. 
The top panels show the predictions of the GP14 model and the bottom panels show the 
predictions for the G13 model. The different colours indicate different redshifts 
as labelled in the top right panel.
In the left panel, the redshift decreases from top to bottom .
In the right panel, the redshift increases from top to bottom.
}
\label{Fig:HOD_z}
\end{figure*}

There are several reasons why the evolution of the HOD is interesting: 
1) it allows us to characterise galaxy evolution, 2) if we can parametrize 
the evolution, the HOD can be used to build mock galaxy catalogues 
that cover a broad range of number densities and redshifts, and 3) observed
clustering measures at different epochs can be modeled consistently.

The choice of using the halo occupation function to quantify the evolution 
in the galaxy population has some distinct advantages over utilizing the 
correlation function. The halo occupation function itself does not depend on 
the distribution of galaxies within a halo, which is something 
different galaxy formation models have modelled in different ways to date 
(see, for example, the discussion in \citealt{C13}). 
Furthermore, the HOD is a function of halo mass, which makes it easier to 
interpret in terms of the implications for galaxy formation models.  
Finally, the parameters of the HOD give fundamental information about the galaxy sample 
(as shown in Sections~\ref{SubSec:FitEv} and \ref{SubSec:M1Min}).
The predictions for the redshift evolution of the HOD have not been widely 
studied over a large redshift range, and such an investigation can inform 
empirical treatments of the evolution (e.g., 
\citealt{Coupon:2012,delaTorre:2013,Skibba:2015}). 

The redshift evolution of the HODs and the correlation functions predicted by 
the G13 and GP14 models are shown in Fig.~\ref{Fig:HOD_z} for galaxy samples 
with a number density of $6.56\times 10^{-3} \, h^{-3}\, {\rm Mpc}$. The 
clustering amplitude increases with time, corresponding to the mean occupation
functions shifting toward larger masses with decreasing redshift. 
This trend is mostly due to the process of hierarchical 
accretion, i.e., the evolution of the halo mass function, coupled with the 
evolution of galaxy bias. 
In fact, the halo mass function exhibits stronger 
evolution and dominates this trend. We also examined the HOD evolution
when plotting the occupation functions against $M_{\rm h}/M_{\rm C}^{*}$, where 
$M_{\rm C}^{*}$ is the characteristic mass\footnote{The characteristic 
halo mass, $M_{\rm C}$,  is defined by $\delta_{\rm C} = \sigma ( M_{\rm C}, z)$, 
where $\delta_{\rm C}$ is the linear theory threshold for collapse at redshift 
$z$ and $\sigma (M)$ is the linear theory variance on a scale that contains 
uncollapsed mass $M$ \citep{Rod-Puebla:2016}.} of the halo mass function. 
We find much weaker evolution of the mass parameters which determine the 
form of the HOD than is experienced by the characteristic mass of the halo 
mass function over the same redshift interval.

The overall evolution of the HOD is similar in the two models. The evolution
of the correlation function shows subtle differences between the models,
particularly on small scales dominated by the distribution of satellite
galaxies. These differences likely arise due to the different treatment of 
satellites in the models (\S~\ref{SubSubSec:SAM}), specifically with 
regard to the distribution and evolution of satellites that have lost their 
subhalo.
These galaxies (known as orphan galaxies) are located in the inner 
part of the halo and contribute to the correlation function on small scales \citep{C13}.  We have examined them in detail for the  
$n = 6.56\times 10^{-3} \, h^{3} \, {\rm Mpc}^{-3}$ case, and find that in
G13 there is a roughly constant fraction of orphan satellite galaxies 
with redshift. 
The fraction is smaller than in GP14, which explains why the correlation function is similar for 
all redshifts on small scales in G13 and is lower in amplitude than GP14.

The qualitative trends are also similar for the other samples with different
number densities.  The amount of information we can obtain from visual 
inspection, however, is limited. To make a more quantitative study, we proceed 
to fit the 5-parameter form to the HODs predicted by the models for the 
different samples and compare the best-fitting values.

\subsection{Fitting the HOD predicted by SAMs}
\label{SubSec:Fit}

To quantify the evolution of the HOD we study the change of the best-fitting 
parameters as a function of redshift and number density. 
The HOD parametrization we use is the five-parameter one presented
in Section~\ref{SubSubSec:Param}.  To make a more 
accurate fit of the model HOD we consider the central and satellite 
galaxies separately, using the classification assigned by the models. 
The fits are carried out using 
a $\chi^2$ minimization method expressed in terms of the logarithm of 
the number density of galaxies. 
We only consider haloes for which the mean occupation satisfies 
$\langle N(M_{\rm h})\rangle > 0.1$. 
This limit was adopted because it is lower than the amplitude of the HOD 
that is typically constrained in observational studies. Also, in this range 
we are not affected by issues which arise from the construction of the merger 
tree used by the SAMs \citep{C13}. Furthermore, the shape of the HOD in this
regime is better described by the 5-parameter form adopted.

When fitting the HOD we weight all mass bins equally. We tested weighting 
the mass 
bins by the number of haloes they contain and by their contribution to the 
effective bias. However, applying weights in these ways tends to over emphasize
a particular part of the HOD leading to considerable discrepancies at high 
halo masses. Instead we treat each mass bin as having the same error. As is 
standard practice in such cases, the uncertainty on the best-fit parameters 
is determined from the $\chi^{2}$ values once normalized to 
$\chi_{\rm min}^{2}/{\rm d.o.f.}=1$. 

\begin{figure}
\includegraphics[width=0.48\textwidth]{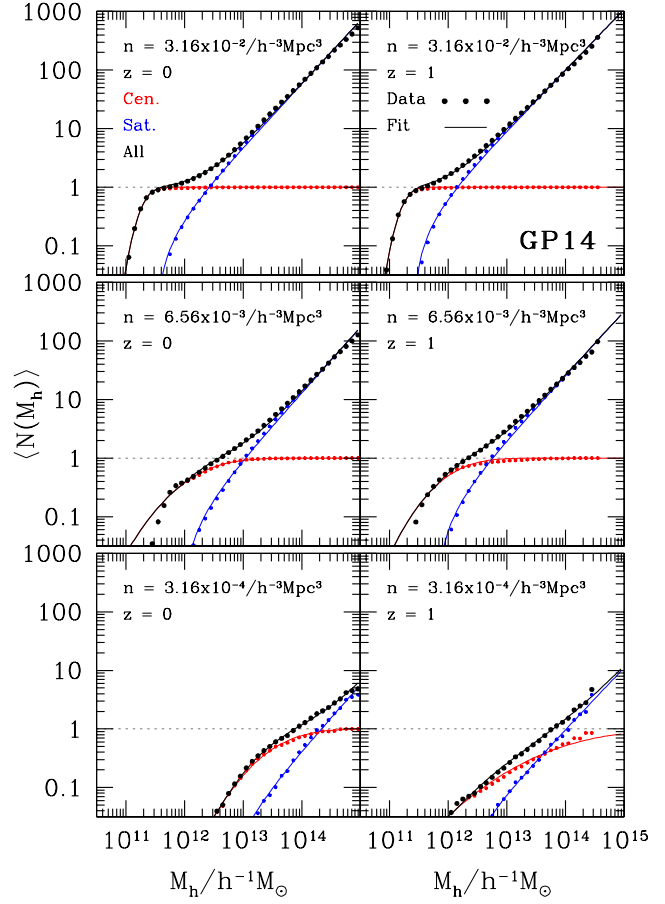}
\caption{
Measured and fitted halo occupation functions for the GP14 model at $z=0$ 
(left panels) and $z=1$ (right panels) for three representative number densities: $ 3.16 \times 10^{-2} \, h^{3} \, {\rm Mpc}^{-3} $ (top), 
$6.56 \times 10^{-3} \, h^{3}\, {\rm Mpc^{-3}}$ (middle) and 
$n =3.16 \times 10^{-4} \, h^{3} \, {\rm Mpc}^{-3}$ (bottom). 
Dots show the HODs predicted by the SAMs and the lines show the 5-parameter
model best fits to them. Black dots and lines represent all galaxies, the
central galaxies are shown in red and the satellites in blue.}
\label{Fig:HOD_Gon}
\end{figure}

\begin{figure}
\includegraphics[width=0.48\textwidth]{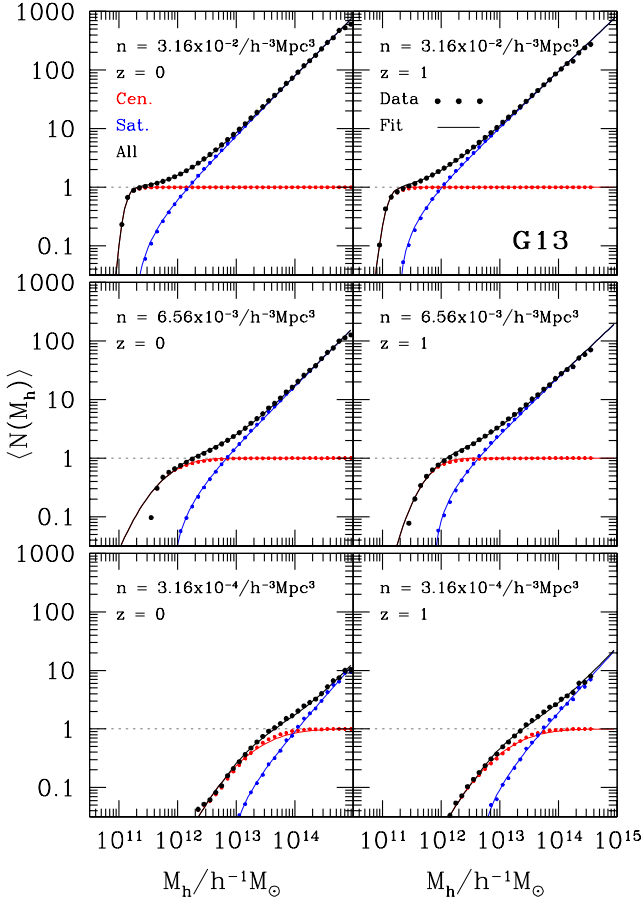}
\caption{
The same as in Fig.~\ref{Fig:HOD_Gon} but for the G13 model. 
}
\label{Fig:HOD_Guo}
\end{figure}

Figs.~\ref{Fig:HOD_Gon} and ~\ref{Fig:HOD_Guo} show the occupation functions
determined from the SAMs together with their best-fitting 5-parameter models 
for GP14 and G13, respectively. The halo occupation functions are shown for
three representative number densities and for $z=0$ and $z=1$.  
The fitted HODs generally produce good fits for all cases over most of the
range.  A deviation is seen in a couple of the cases at the very low mass end,
related to fitting only above $\langle N(M_{\rm h})\rangle > 0.1$ (however 
lowering this limit in an attempt to remedy these discrepancies generally resulted in a 
worse fit in the turnover of the central occupation function).

The predicted occupation functions look very similar in the two models for 
the sample with the highest number density of galaxies (top panels). There 
is a clear plateau in the total halo occupation due to 
central galaxies until the halo mass at which the satellites power-law 
starts to dominate. For the intermediate number density sample, in the GP14 
model the central HOD reaches unity around the same mass that the satellite 
occupation does. In contrast, for the G13 model the central occupation 
reaches unity at a lower halo mass than the satellite occupation, which 
results in more of a step-like shape for the overall HOD. 
This reflects differences in the treatment of the suppression of cooling 
by active galactic nuclei in the two models. For the lowest number density 
sample, both models display a very broad turnover in the central occupations  
and no distinct plateau. The central galaxy HOD does reach unity for this 
sample (though only just for the GP14 model at $z=1$). 

Some subtle general differences with redshift are noticeable for all number 
density samples and both models. Going toward higher redshift, the plateau 
in the mean halo occupation decreases somewhat and the width of the central
occupation appears to change. We examine the evolutionary changes in detail 
in \S~\ref{SubSec:FitEv}.

\subsection{Modelling the HOD evolution}
\label{SubSec:FitEv}

\begin{figure}
\includegraphics[width=0.48\textwidth]{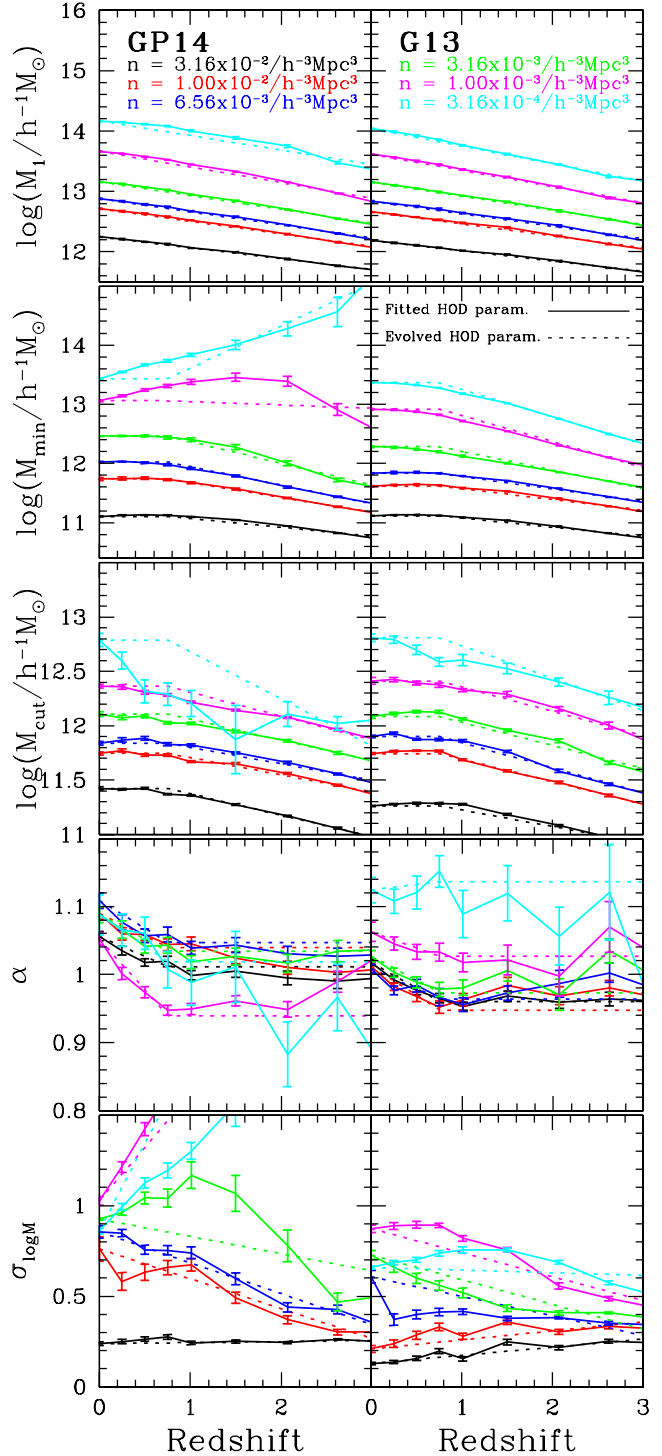}
\caption{
The evolution with redshift of the 5 HOD parameters, after fitting to the 
predictions of the GP14 (left) and G13 (right) SAMs. 
From top to bottom the properties shown in each row are: 
$M_1$, $M_{\rm min},M_{\rm cut}$, $\alpha$ and $\sigma_{\log M}$. 
The different colors represent different number densities as 
labelled in the top panels. 
Error bars represent the standard deviation from the fitted parameter value 
(as explained in the text). Dotted lines represent the fits of Eqs.\ 5 to 9.
In the three top panels, the number density increases from top to bottom. 
In the two bottom panels,the number density tends to increase from top to 
bottom, except for the highest number densities (where 
the HOD is not well defined).
}
\label{Fig:Param_Ev}
\end{figure}

To quantify the evolution of the HOD we focus on how the best-fitting 
parameters change with redshift.  Fig.~\ref{Fig:Param_Ev} shows the values of 
the best-fitting parameters for the full range of redshifts and number 
densities studied (solid lines connecting points with errorbars).

Fig.~\ref{Fig:Param_Ev} allows us to assess which evolutionary behaviour in 
the HOD parameters is generic and independent of the modelling choices and 
assumptions. The halo mass for which there is typically one satellite galaxy 
per halo, $M_{1}$, evolves in a remarkably similar way in both models, 
declining by 0.6 dex between $z=0$ and $z=3$. This is much weaker than the 
evolution expected in the characteristic halo mass which changes 
by four orders of magnitude over the same redshift interval 
(\citealt{Rod-Puebla:2016}). 
The evolution of $M_{\rm min}$, the minimum halo mass at which half the haloes 
host a central evolves in a similar way between the models for the four 
highest density galaxy samples. For the two samples with the lowest space 
densities of galaxies, $M_{\rm min}$ increases with redshift in the GP14 model, 
but continues to decline with increasing redshift in G13. Globally (with the
exception of the two least abundant samples in GP14),
the evolution seems more modest for $M_{\rm min}$ than we found for $M_{1}$: 
$M_{\rm min}$ is roughly constant to $z \sim 0.75$ before dropping by 0.2-0.6 dex 
depending on the abundance of the sample. $M_{\rm cut}$, the cutoff mass for
hosting a satellite galaxy,  evolves in a similar way in G13 and GP14, though 
the results for the least abundant sample in GP14 have large errors. 
The slope of the satellite HOD power laws are different between the models 
but show little dependence on redshift. The largest differences are found in 
the evolution of the width of the transition from zero to unity in the central 
galaxy HOD. We have checked the evolution of the parameters also with earlier
SAM catalogues from the two groups and generally find a similar behaviour to 
that shown by the models studied here.

To quantify the evolution of the best-fitting HOD parameters we use a 
single evolutionary parameter for each property, $\gamma$, 
along with the value of the parameter at $z=0$. This approach will allow us 
to specify the value of the parameter at redshifts intermediate to the ones 
where we have SAM outputs, which will be important for building mock catalogues,
as well as provide a generalized form of the HOD as a function of redshift.
We fit these for each of our number density samples independently using the values
of the fitting parameters shown in Fig.~\ref{Fig:Param_Ev}. We 
represent the value of $M_1$ as a function of redshift as a power law:
\begin{equation}
 \log M_{1}(z) =  \log M_{1}(z=0) + \gamma_{M_{1}} \times z.
\end{equation}
For $M_{\rm min}$ and $M_{\rm cut}$ we use a constant value from $z=0$ to
$0.75$, followed by a power law;
\begin{equation}
\log M_{\rm min}(z) = 
 \left\{
	\begin{array}{ll}
		\log M_{\rm min}(z=0) & \mbox{if } z \leq 0.75\\
		\log M_{\rm min}(z=0) + (z-0.75) \times \gamma_{M_{\rm min}}  & \mbox{if } z > 0.75,
	\end{array}
\right.
\end{equation}
and
\begin{equation}
\log M_{\rm cut}(z) = 
 \left\{
	\begin{array}{ll}
		\log M_{\rm cut}(z=0)) & \mbox{if }  z \leq 0.75\\
		\log M_{\rm cut}(z=0) + (z-0.75) \times \gamma_{M_{\rm cut}}  & \mbox{if } z > 0.75.
	\end{array}
\right.
\end{equation}
For $\alpha$ we use a power law value from z=0 to 0.75, 
followed by a constant value;
\begin{equation}
\alpha(z) = 
 \left\{
	\begin{array}{ll}
		\alpha (z=0) + z \times \gamma_{\alpha} & \mbox{if }  z \leq 0.75\\
		\alpha (z=0) + 0.75 \times \gamma_{\alpha}  & \mbox{if } z > 0.75.
	\end{array}
\right.
\end{equation}
Finally, for $\sigma_{\log M}$ we model the evolution as linear in redshift
\begin{equation}
\sigma_{\log M}(z) =  \sigma_{\log M}(z=0) + \gamma_{\sigma_{\log M}} \times z.
\end{equation}
Note that this is a first order approximation, since we were not able to find 
a simple form that describes the evolution of this parameter. 

The fits to the evolution of the HOD parameters are shown by the dotted lines 
in Fig.~\ref{Fig:Param_Ev}. The values of the parameters at $z=0$ are presented 
in Tables 1 \& 2, while the values for the evolutionary parameters 
($\gamma$) are shown in Tables 3 \& 4, 
for GP14 and G13 respectively. 

\begin{table}
\caption{The values of $M_1$, $M_{\rm min}$, $M_{\rm cut}$, $\alpha$ and 
$\sigma_{\log M}$
for the GP14 model at $z=0$, shown for the 6 fixed number density samples.
We also provide the corresponding stellar mass thresholds, $M_{*}^{\rm thres}$, 
for these samples.
All masses are in units of $h^{-1} M_{\odot}$.}
\begin{center}
\begin{tabular}{ |c|c|c|c|c|c|c| }
\hline
$ n/h^{-3} {\rm Mpc}^3$  &  $\log M_{*}^{\rm thres}$  &  $\log M_{1}$  &   $\log M_{\rm min}$  &   $\log M_{\rm cut}$  &  $\alpha$  &  $\sigma_{\log M}$ \\ \hline
$3.16\times10^{-4}$  &  11.02  &  14.17  &  13.43  & 12.79  &  1.09  &  0.86\\ 
$1.00\times10^{-3}$  &  10.86  &  13.66  & 13.06  &  12.36  &  1.05  &  1.03\\ 
$3.16\times10^{-3}$  &  10.53  &  13.16  & 12.46  &  12.11  &  1.09  &  0.93\\ 
$6.56\times10^{-3}$  &  10.14  &  12.88  & 12.02  &  11.84  &  1.11  &  0.86\\ 
$1.00\times10^{-2}$  &  9.90  &  12.71  &  11.74  &  11.75  &  1.09  &  0.76\\ 
$3.16\times10^{-2}$  &  9.07  &  12.25  &  11.11  &  11.42  &  1.06  &  0.24\\ 
\hline
 \end{tabular}
\end{center}
\label{tab:Gonzalezz0Val}
\end{table}

\begin{table}
\caption{
Same as Table 1 (parameter values at $z=0$) but for the G13 model.}
\begin{center}
\begin{tabular}{ |c|c|c|c|c|c|c| }
\hline
$ n/h^{-3} {\rm Mpc}^3$  &  $\log M_{*}^{\rm thres}$  &  $\log M_{1}$  &  $\log M_{\rm min}$ &   $\log M_{\rm cut}$  &  $\alpha$   &  $\sigma_{\log M}$ \\ \hline
$3.16\times10^{-4}$  &  10.96  &  14.04  &  13.36  &  12.81  &  1.12  &  0.66 \\ 
$1.00\times10^{-3}$  &  10.77  &  13.61  &  12.92  &  12.41  &  1.06  &  0.87 \\ 
$3.16\times10^{-3}$  &  10.53  &  13.15  &  12.28  &  12.08  &  1.03  &  0.73 \\ 
$6.56\times10^{-3}$  &  10.29  &  12.84  &  11.83  &  11.90  &  1.01  &  0.61 \\ 
$1.00\times10^{-2}$  &  10.10  &  12.66  &  11.62  &  11.74  &  1.01  &  0.21 \\ 
$3.16\times10^{-2}$  &  9.21  &  12.19  &  11.12  &  11.26  &  1.02  &  0.13 \\ 
\hline
 \end{tabular}
\end{center}
\label{tab:Guo13z0Val}
\end{table}

\begin{table}
\caption{The values of the evolution parameters $\gamma$ for 
$M_1$, $M_{\rm min}$, $M_{\rm cut}$, $\alpha$ and $\sigma_{\log M}$
for the GP14 model, shown for the 6 fixed number density samples.
}
\begin{center}
\begin{tabular}{ |c|c|c|c|c|c| }
\hline
$ n/ h^{-3} {\rm Mpc}^3$  &  $\gamma_{\log M_{1}}$  &  $\gamma_{log M_{\rm min}}$ &  $\gamma_{\log M_{\rm cut}}$  &  $\gamma_{\alpha}$  &  $\gamma_{\sigma_{\log M}}$ \\ \hline
$3.16\times10^{-4}$  &  -0.24  &  0.49  &  -0.43  &  -0.09  &  0.96 \\ 
$1.00\times10^{-3}$  &  -0.26  &  -0.01 &  -0.22  &  -0.15  &  0.59 \\ 
$3.16\times10^{-3}$  &  -0.23  &  -0.24  &  -0.19  &  -0.07  &  -0.10 \\ 
$6.56\times10^{-3}$  &  -0.22  &  -0.21  &  -0.15  &  -0.08  &  -0.17 \\ 
$1.00\times10^{-2}$  &  -0.21  &  -0.17  &  -0.16  &  -0.07  &  -0.16 \\ 
$3.16\times10^{-2}$  &  -0.18  &  -0.10  &  -0.20  &  -0.06  &  0.01 \\ 

\hline
 \end{tabular}
\end{center}
\label{tab:GonzalezEvParam}
\end{table}

\begin{table}
\caption{Same as Table 3 (values of evolution parameters) but for the 
G13 model.}
\begin{center}
\begin{tabular}{ |c|c|c|c|c|c| }
\hline
$ n/h^{-3} {\rm Mpc}^3$  &  $\gamma_{\log M_{1}}$  &  $\gamma_{log M_{\rm min}}$  &  $\gamma_{\log M_{\rm cut}}$  &  $\gamma_{\alpha}$   &  $\gamma_{\sigma_{\log M}}$ \\ \hline
$3.16\times10^{-4}$  &  -0.29  &  -0.31  &  -0.30  &  0.02  &  -0.01 \\ 
$1.00\times10^{-3}$  &  -0.27  &  -0.30  &  -0.22  &  -0.05  &  -0.13 \\ 
$3.16\times10^{-3}$  &  -0.23  &  -0.21  &  -0.21  &  -0.07  &  -0.14 \\ 
$6.56\times10^{-3}$  &  -0.21  &  -0.14  &  -0.23  &  -0.06  &  -0.11 \\ 
$1.00\times10^{-2}$  &  -0.20  &  -0.12  &  -0.21  &  -0.09  &  0.05 \\ 
$3.16\times10^{-2}$  &  -0.17  &  -0.10  &  -0.15  &  -0.08  &  0.04 \\ 
\hline
 \end{tabular}
\end{center}
\label{tab:Guo13EvParam}
\end{table}

\begin{figure}
\includegraphics[width=0.48\textwidth]{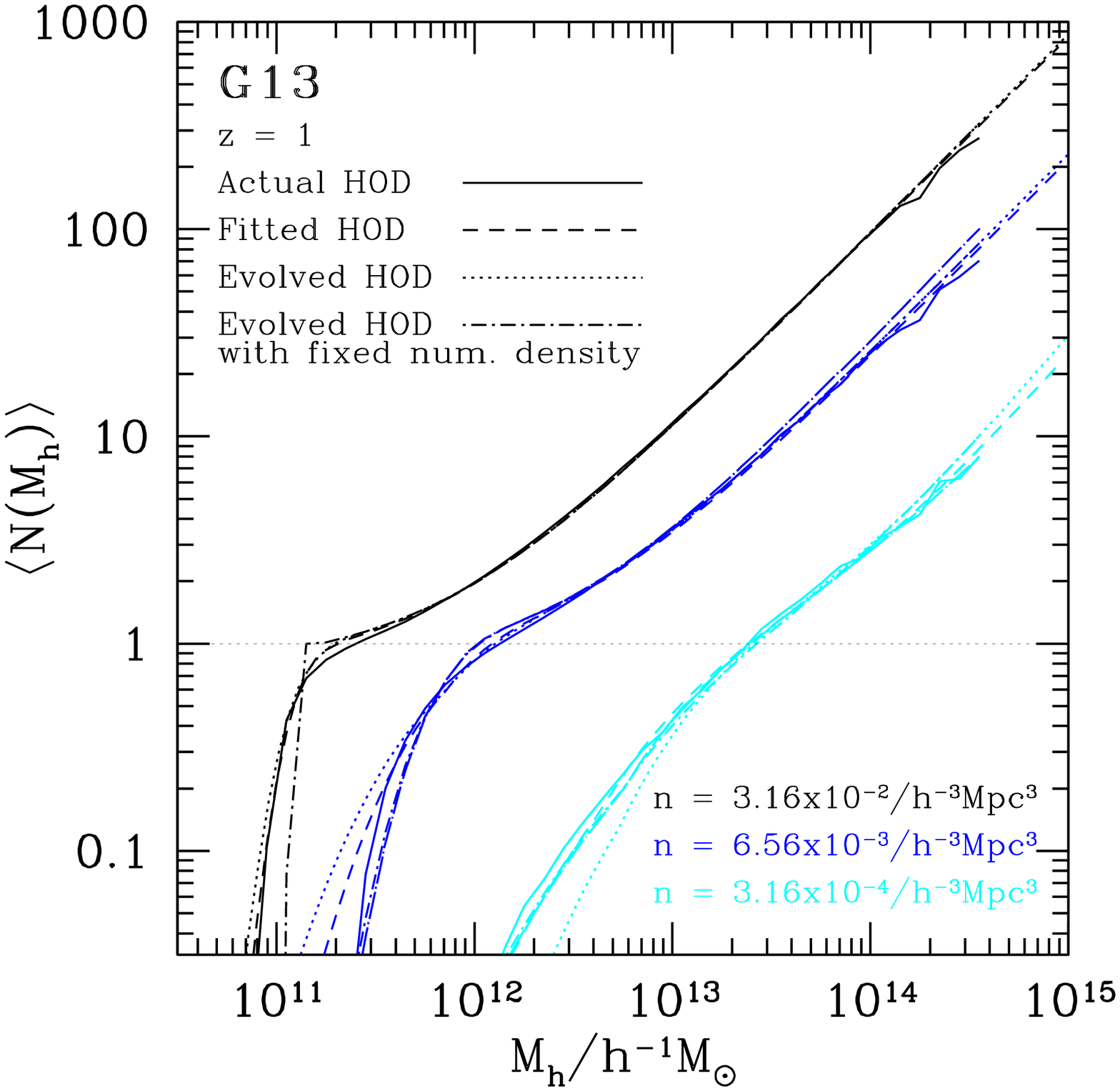}
\includegraphics[width=0.48\textwidth]{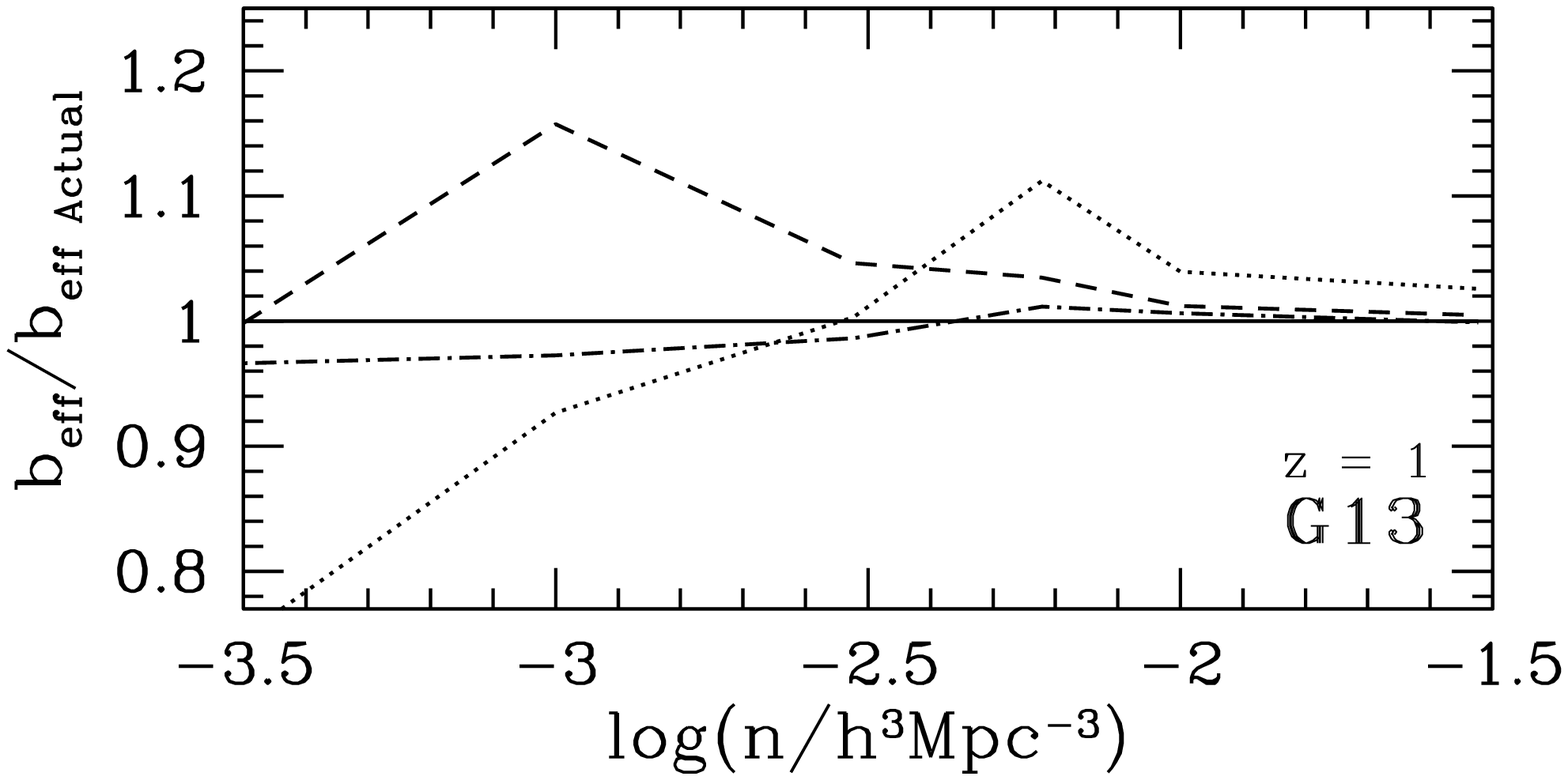}
\caption{
(Top) The halo occupation functions for the G13 model at $z=1$, for three number 
densities as labelled. The ``Actual HOD'' results, i.e. the direct output in 
the simulation, are shown as solid lines. ``Fitted HOD'' (dashed lines) 
correspond to the 5-parameter fits to the ``Actual HOD'' at $z=1$. The 
``Evolved HOD'' (dotted lines) is obtained from our fits to the redshift
dependence of each parameter, evolved from their value at $z=0$.
Finally, ``Evolved HOD with fixed number density'' 
(dot-dashed lines) shows the HOD assuming the redshift dependence for the 
fit parameters, but excluding $\sigma_{\log M}$, which instead is determined
by requiring that the HOD fit reproduces the number density of the sample.  
The number density decreases from top to bottom. (Bottom) Effective 
bias ($b_{eff}$) calculated for the ``Fitted HOD'', ``Evolved HOD''  and 
``Evolved HOD with fixed number density'' cases plotted relative to that of 
the ``Actual HOD'' as function of number density, for the G13 model at z = 1.
(See text for more details.) }
\label{Fig:HOD_Parm}
\end{figure}

We have also compared the evolution of the fitting parameters  $M_1$ and 
$M_{min}$ with their exact values extracted from the HOD (i.e., the halo mass 
of which $\langle N_{\rm sat}(M = M_1)\rangle =  1$ and 
$\langle N_{\rm cen}(M = M_{\rm min})\rangle = 0.5$). We find good agreement 
between these values and those obtained by fitting the HOD. Also, their 
redshift evolution is consistent with the models proposed
in this work. We show that at least for these two parameters, the evolution 
is well constrained. Thus, we do not expect that any potential degeneracy in 
the fitting of the parameters would affect the evolutionary model we propose.

To test the accuracy of this approximation for the evolution of the HOD 
parameters, we plot in Fig.~\ref{Fig:HOD_Parm} the occupation functions
obtained using the parameter values derived from our fits for the redshift 
dependence (eq.~5-9; labelled ``Evolved HOD'' in the plot)
and compare these with the occupation functions predicted by the SAM 
(labelled ``Actual HOD''). 
We do this for G13 at $z=1$ and for three different number densities: 
$3.16\times 10^{-4}\, h^{3}\, {\rm Mpc}^{-3}$ (cyan lines), 
$6.56\times 10^{-3} \, h^{3} \, {\rm Mpc}^{-3}$ (blue lines) and 
$3.16\times 10^{-2}\, h^{3} \,{\rm Mpc}^{-3}$ (black lines). 
We have used G13 for this exercise since the HODs predicted by this model 
are better described by the 5-parameter fit over 
a wider range of number densities and redshifts than is the case for the GP14 model. 

Fig.~\ref{Fig:HOD_Parm} shows the halo occupations obtained from the parameter
evolution fits is a reasonably good match to the direct output by the models 
at all number 
densities. The main differences are found at lower halo masses and are 
caused by the limitations in fitting $\sigma_{\log M}$.  An alternative 
way of modeling the parameter evolution is shown by the dash-dotted lines (labelled ``Evolved HOD with fixed number density'' in the plot).
In this case, instead of fitting $\sigma_{\log M}$, we have set the other 
parameters to the values given by the evolutionary fit, and we fix the 
value of $\sigma_{\log M}$ to reproduce the number density of the sample.  
The resulting HOD gives a better reproduction of the model HOD for the 
lower number density samples. At the highest number density, the values 
of the evolved parameters  overestimate the number density. To compensate for 
this, $\sigma_{LogM}$ takes the minimum allowed value ($\sigma_{LogM} \sim 0$).

To investigate the significance of these deviations, we calculate the
effective bias of the predicted G13 HODs at z=1 , following the procedure of 
\cite{Kim:2009}. We show the ratio of the different effective biases to that
of the actual HOD in  Fig.~\ref{Fig:HOD_Parm}. The ``Evolved HOD''  model 
shows small differences in the effective bias ($< 4\%$) for the highest number 
density, while the lowest number density has a considerable difference of 
$\sim 25\%$. In the case of ``Evolved HOD with fixed number density'', we 
find differences $\lesssim 3\%$ for all number densities. This means that the 
first method can reliably reproduce the clustering signal at high number 
densities, while the second method can do so for a broader range of number 
densities. Interestingly, just fitting the HOD with the 5-parameter model 
(labelled as ``Fitted HOD'' in the figure) can by itself produce differences 
of over 10\% in the effective bias, due to the limitation of the accuracy with 
which this form can fit  the detailed distribution.

We also test how the HOD evolves if $\alpha$ is kept constant. By 
evolving the HOD to z=1, we find minimal differences in the HODs shown in 
Fig.~\ref{Fig:HOD_Parm}. We thus note that the evolution of $\alpha$ has a 
minor impact on the evolution of the HOD at low redshifts.

Other approximations for the evolution of the HOD are mentioned in the 
literature. \citet{delaTorre:2013} approximate jointly the HOD dependence on 
luminosity and redshift from VIPERS clustering measurements. 
\citet{Manera:2015} incorporates a simplified evolving HOD to SDSS-III mock 
catalogues based on a compilation of HOD measurements \citep{Parejko:2013}.
\citet{Hearin:2016} model the stellar-to-halo-mass relation  based on 
abundance matching predictions \citep{Behroozi:2010}.
Direct comparison with these is not straightforward, however, due to 
differences in the sample definitions, HOD forms, approaches, and assumptions 
made.

\subsection{Evolution of the $M_1/M_{\rm min}$ ratio}
\label{SubSec:M1Min}

One relation that is often extracted from the HOD is the ratio between the
two characteristic halo masses, 
$M_{1}/M_{\rm min}$ 
\citep{Zehavi:2011,Coupon:2012,Guo:2014b,Skibba:2015}. 
This ratio links the mass at which haloes start being populated
by central galaxies (specifically, where 
$\langle N_{\rm cen}(M_{\rm min})\rangle = 0.5$)
and the mass at which the halo starts hosting satellites as well
(i.e., the halo mass for which $\langle N_{\rm sat}(M_1)\rangle = 1$). 
Larger values of $M_1/M_{\rm min}$ indicate that central galaxies populate haloes 
over a broader range of halo masses before satellite galaxies start to 
dominate, resulting in the ``plateau'' feature in the HOD (as seen, for 
example, in the high density galaxy samples in Figs.~\ref{Fig:HOD_Gon} 
and ~\ref{Fig:HOD_Guo}). 
Haloes in the ``hosting gap'' mass range between $M_{\rm min}$ and $M_1$ tend to
host more massive central galaxies rather than multiple galaxies 
\citep{Berlind:2003}.
The exact value of the ratio reflects the balance between accretion and
destruction of the satellites \citep{Zentner:2005,Watson:2011} and
the ratio also has a strong influence on the shape
of the correlation function (e.g., \citealt{Seo:2008,Watson:2011,Skibba:2015}).

\begin{figure}
\includegraphics[width=0.48\textwidth]{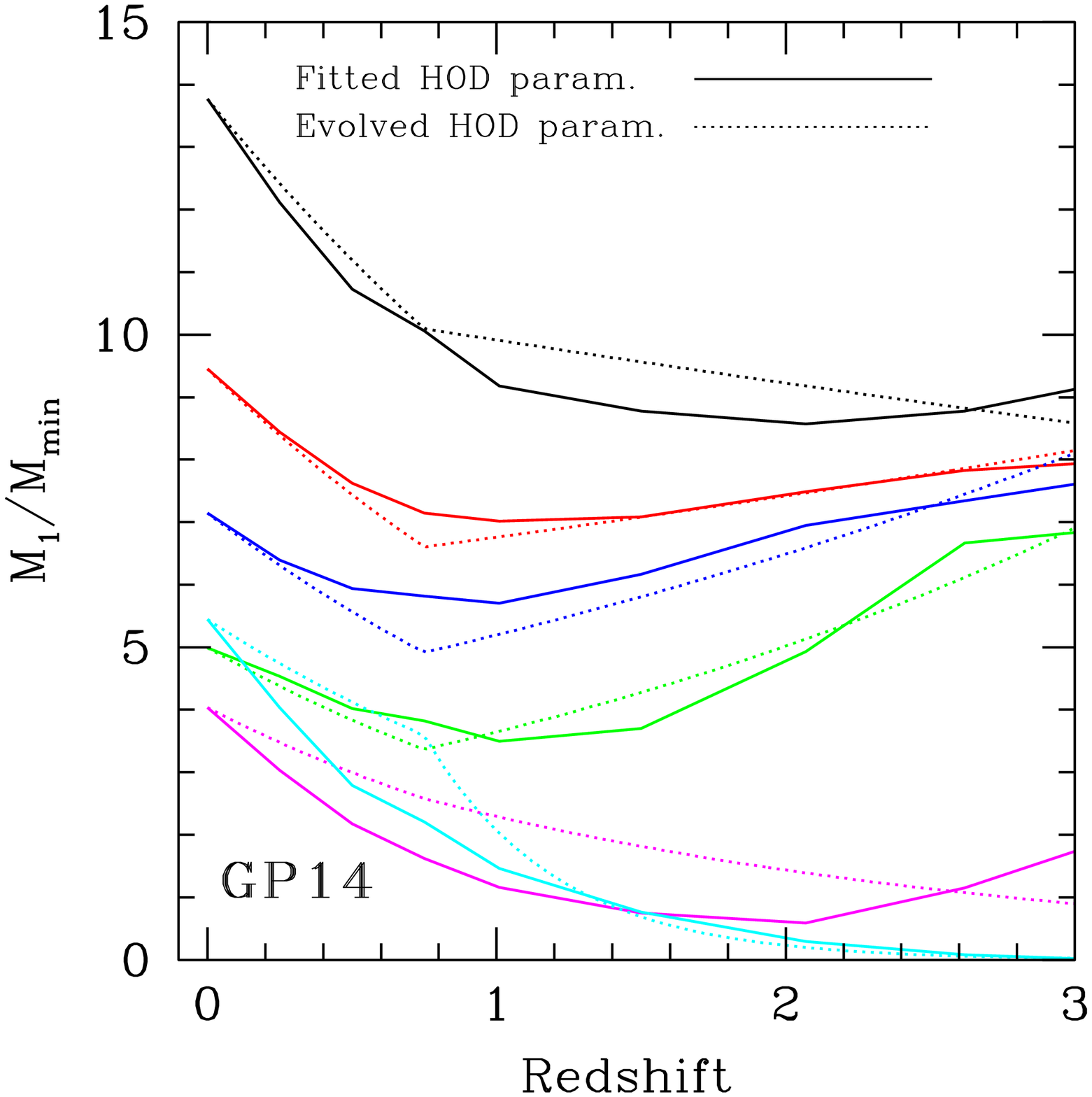}
\includegraphics[width=0.48\textwidth]{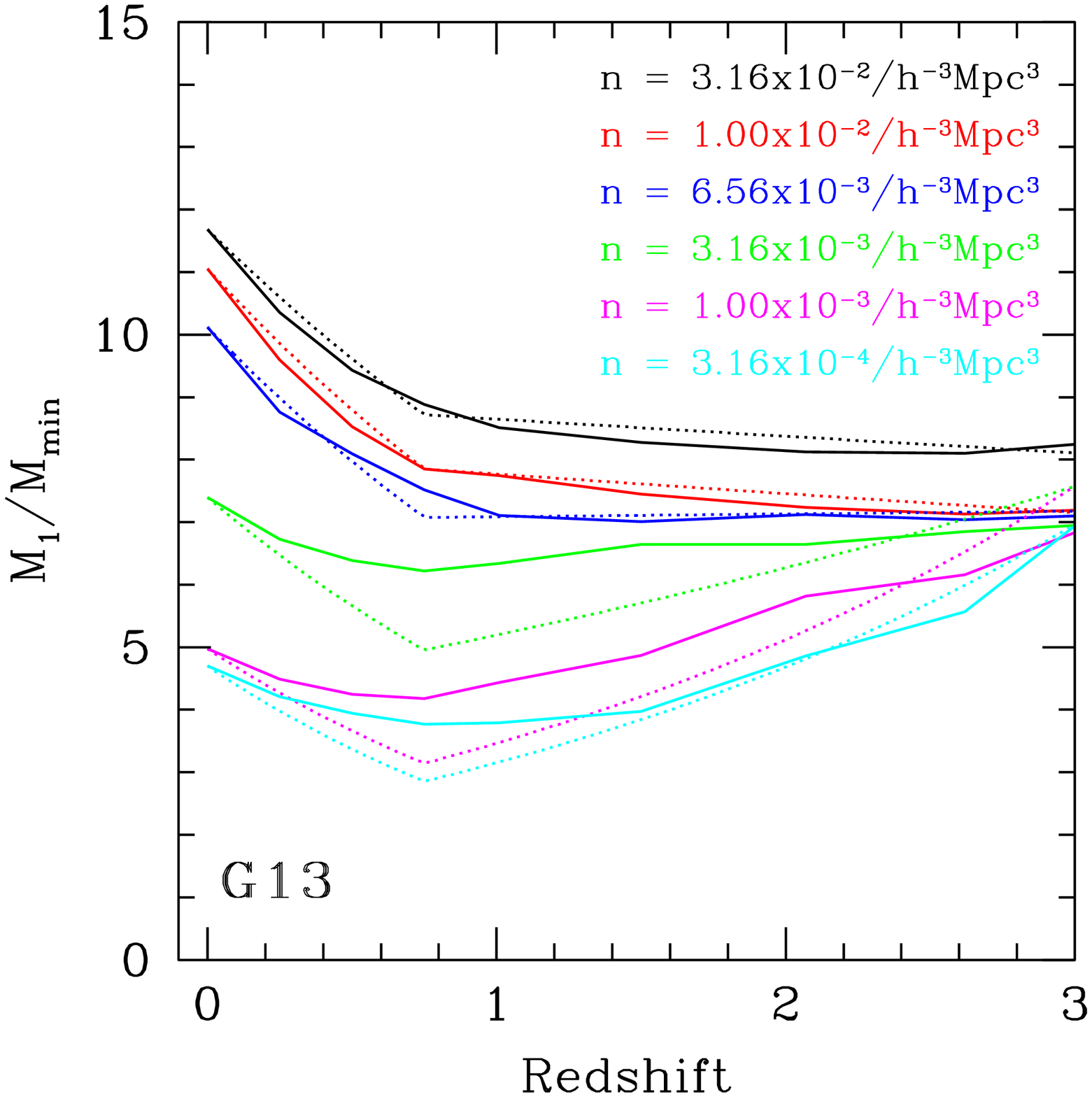}
\caption{
The ratio $M_1$/$M_{\rm min}$ plotted as a function of redshift for the 
GP14 (top) and G13 (bottom) panels. Different colors indicate different
number densities as  labelled. Solid lines show the ratio obtained from 
the best-fitting parameters to the HOD output by the models. Dashed lines 
show the ratio using the best-fitting redshift evolution using Eqs. 5 and 6.
The number density decreases from top to bottom (with the exception of
the lowest number density sample, which crosses the adjacent sample for
$z<1.5$ in the top panel). 
}
\label{Fig:Ratio}
\end{figure}

Fig.~\ref{Fig:Ratio} shows the evolution of $M_1/M_{\rm min}$ for the GP14 
and G13 models, for our different number density samples.
The solid line shows the ratio between $M_{1}/M_{\rm min}$
obtained by fitting the HOD using Eqs.~1 and 3, while the dotted line shows the 
prediction from the evolutionary model presented in the previous section 
(Eqs.~5 and 6). The evolution of this ratio with redshift is complex. Its 
shape is different for each number  density, and cannot be described by a 
simple functional form, though it is reassuring that our simple evolution
model for $M_{\rm min}$ and $M_1$ also captures reasonably well the
behavior of their ratio.

We note that the value of $M_1/M_{\rm min}$ increases as we move to higher 
number densities (for any fixed redshift). This is in agreement with the 
results derived from observations (see, e.g., Fig.~4 of \citealt{Guo:2014b}).
Decreasing the number density corresponds to more massive galaxies, which 
reside in more massive haloes, as we saw in Fig.~\ref{Fig:n_ev}. The trend
we see likely reflects the relatively late formation of these massive haloes,
which leaves less time for satellites to merge onto central galaxies and
thus lowers the satellite threshold $M_1$ and this ratio.

As far as the redshift evolution, the ratio $M_1/M_{\rm min}$ decreases with 
redshift until $z \sim 0.75$ and then stays constant or increases (for all 
but the lowest number densities). The decrease for moderately increasing
redshifts is probably due to a similar reasoning: for increasing redshift there is
less time for destruction of the satellites resulting in a smaller ratio.
For higher redshifts, $M_{\rm min}$ evolves as well and the trend halts or
reverses and the exact behavior is more complex to predict.

There are clear differences between the models we study in terms of the
range of values of this ratio.  This difference is perhaps related to
the different treatment of satellites. 
Measurements of $M_1/M_{\rm min}$ and its evolution may thus provide strong
constraints on models of galaxy formation.

Given the clear (yet complex) evolutionary trends present, we caution the
reader against assuming an overall constant shift in halo mass of the halo 
occupation functions with redshift, corresponding to a constant ratio of
$M_1/M_{\rm min}$.
The broad sense of lower values for this ratio when going toward higher 
redshift is also consistent with predictions of abundance matching modeling in
dissipationless simulations \citep{Kravtsov:2004,Conroy:2006} as well as 
with inferred values from observations of galaxy clustering
(e.g., \citealt{Zheng:2007,Coupon:2012,Skibba:2015}).

\section{Comparison with empirical methods to describe the evolving galaxy population}
\label{Sec:PC2}

We now compare the evolution of the HOD predicted by SAMs with alternative
heuristic approaches  that are sometimes used in the literature to describe 
the evolution of the galaxy population. We focus on the evolution of the
$M_1/M_{\rm min}$ ratio, that we already saw can provide important insight to
galaxy formation and evolution, and on the change in the fraction of galaxies 
that are satellites, $f_{\rm sat}$.

\subsection{Evolution models}
\label{SubSec:EvolModels}

In this discussion, all methods are defined in reference to an N-body 
simulation which follows the evolution of clustering of the dark matter. 
The models we compare, along with the labels used to refer to these models 
in the subsequent discussion, are:

\textbf{Fixed number density:}  This is the model discussed in the previous 
sections, using the output from SAMs. At each redshift, samples are 
constructed by ranking galaxies in order of descending stellar mass. Galaxies 
are retained down to the stellar mass which allows the sample to attain the 
desired number density. This procedure is repeated anew at each redshift, 
without any consideration of the galaxies included in samples at other 
redshifts. There can be considerable churn in the galaxies which make up a 
sample defined by a fixed number density at different redshifts. Galaxies 
may merge, and so no longer exist as a distinct entity at a subsequent 
redshift. Differences in star formation rates between galaxies mean that 
some galaxies may not gain stellar mass as quickly as others and so may lose 
their place on the list of galaxies that make up the sample at a later 
redshift, being replaced by a galaxy that was not previously included. So, 
although the number density of the sample does not change with redshift, the 
membership of the sample is not fixed (see further discussion in 
\S~\ref{SubSec:Comparison}).

\textbf{Tracking evolution:} The tracking model follows the same galaxy
population across time. 
In this case, the galaxy samples are defined at a specified redshift as 
described above, by ranking in order of decreasing stellar mass and retaining 
all the galaxies down to a particular mass to achieve a given number density. 
This exact sample of galaxies is then followed using the SAM implemented in 
the N-body simulation. The size of the sample can shrink as galaxies merge 
according to the treatment of galaxy mergers in the SAM. Note that at a 
subsequent redshift, the sample of galaxies in the tracking evolution model 
can differ substantially from the fixed number density sample outlined above. 
This is because the galaxies in the tracking evolution case are not necessarily 
the most massive at a redshift subsequent to the one at which the sample is 
defined. Different tracking evolution samples can be defined by changing the 
redshift at which the sample is initially specified. Star formation is 
effectively ignored after the redshift at which the sample is defined since 
the sample membership is not reconsidered, but it provides a somewhat idealized
way probing galaxy evolution by tracking an identical set of galaxies over time.

\textbf{Passive evolution:} The passive evolution model imposes strict 
assumptions regarding the physical ``passivity'' of the galaxies, following
an unchanged galaxy population. The starting point is again the output of 
the SAM in the N-body simulation at a specified redshift, selecting all 
galaxies above a specific stellar mass to reproduce a set number density. 
The passive evolution model differs from the tracking evolution model in 
that the number of objects is preserved. If, according to the SAM, two 
galaxies merge, the remnant galaxy counts twice in the HOD, effectively 
doubling the weight of the remnant in any clustering prediction. Again, 
star formation is ignored after the redshift at which the sample is 
defined. In such a passive evolution each galaxy keeps its own identity and
there is no merging or disruption of satellites or formation of new ones. Such 
strong assumptions lend themselves to theoretical predictions of the evolution 
and empirical comparisons (e.g., \citealt{Fry:1996,White:2007,Guo:2014,Skibba:2014}). A detailed study of the evolution of clustering and the HOD under 
passive evolution was presented by \cite{Seo:2008}. 

\textbf{Descendant clustering selection:} This method was proposed by 
\cite{Padilla:2010} to investigate the clustering of the descendants of a 
population of galaxies observed at $z>0$. 
As before, our starting point is 
the output of the SAM model at a particular redshift. The aim is to select 
a sample of dark matter haloes that has the same clustering as the galaxy 
sample: in the original method the galaxy sample in question was an 
observational sample, here it is the output of the SAM ranked by stellar mass. 
The clustering of the galaxy sample is characterised in terms of the median 
host halo mass. A sample of dark matter haloes is then constructed with the 
same median mass, starting from the most massive haloes in the simulation 
and giving each halo equal weight 
(so effectively $ \langle N(M_h) \rangle = 1$). 
These haloes are then followed in the simulation and their evolved median 
halo mass
is used to identify descendants of the original sample (again ranked by stellar
mass). The underlying assumption is that no 
objects enter or leave the sample between the selection redshift and the 
redshift at which the descendants are considered but mergers can take place.
The number of descendant 
haloes can be smaller than the number at the selection redshift following 
mergers between haloes. 

\subsection{Comparison of results}
\label{SubSec:Comparison}

\begin{figure}
\includegraphics[width=0.48\textwidth]{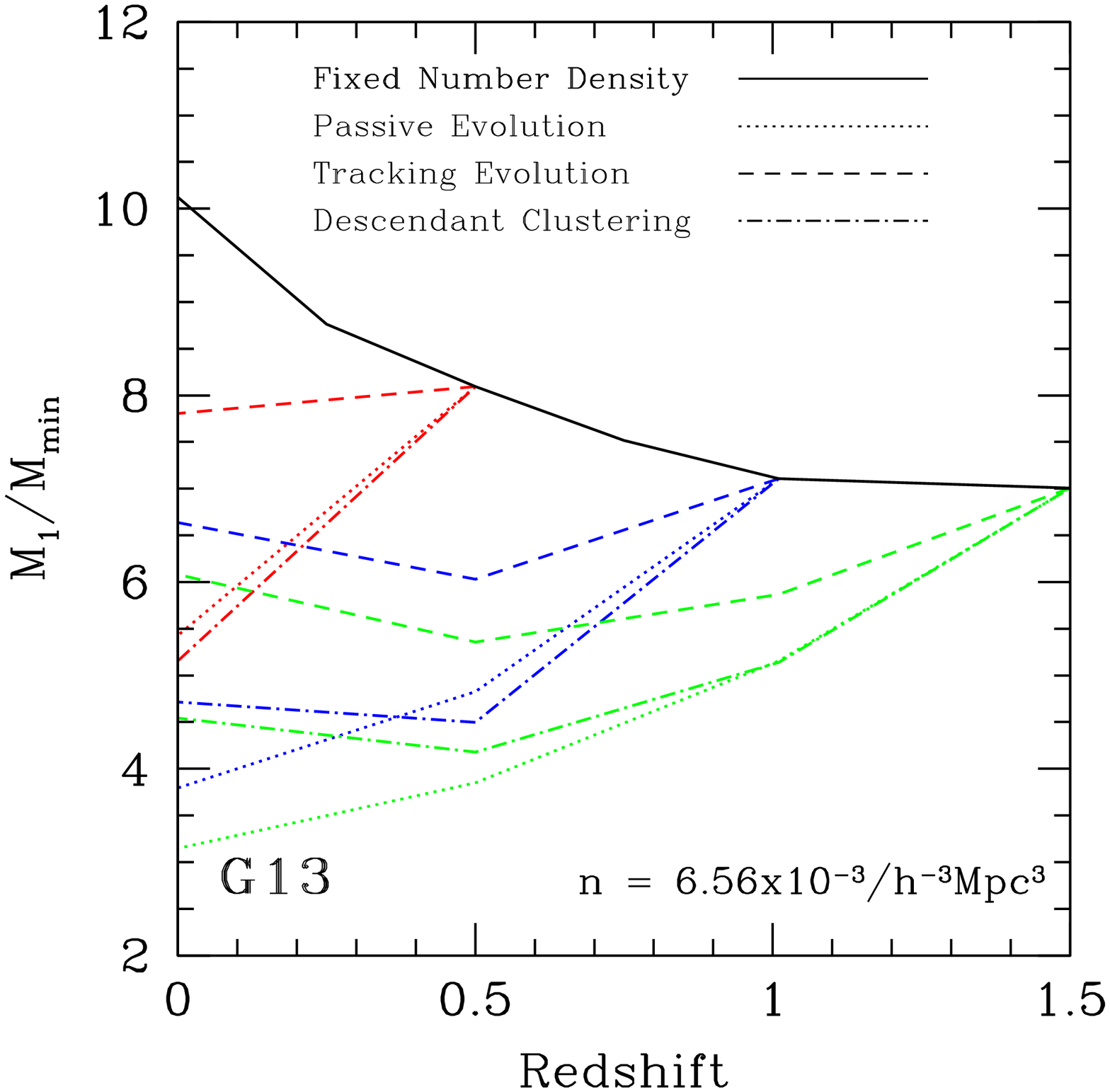}
\includegraphics[width=0.48\textwidth]{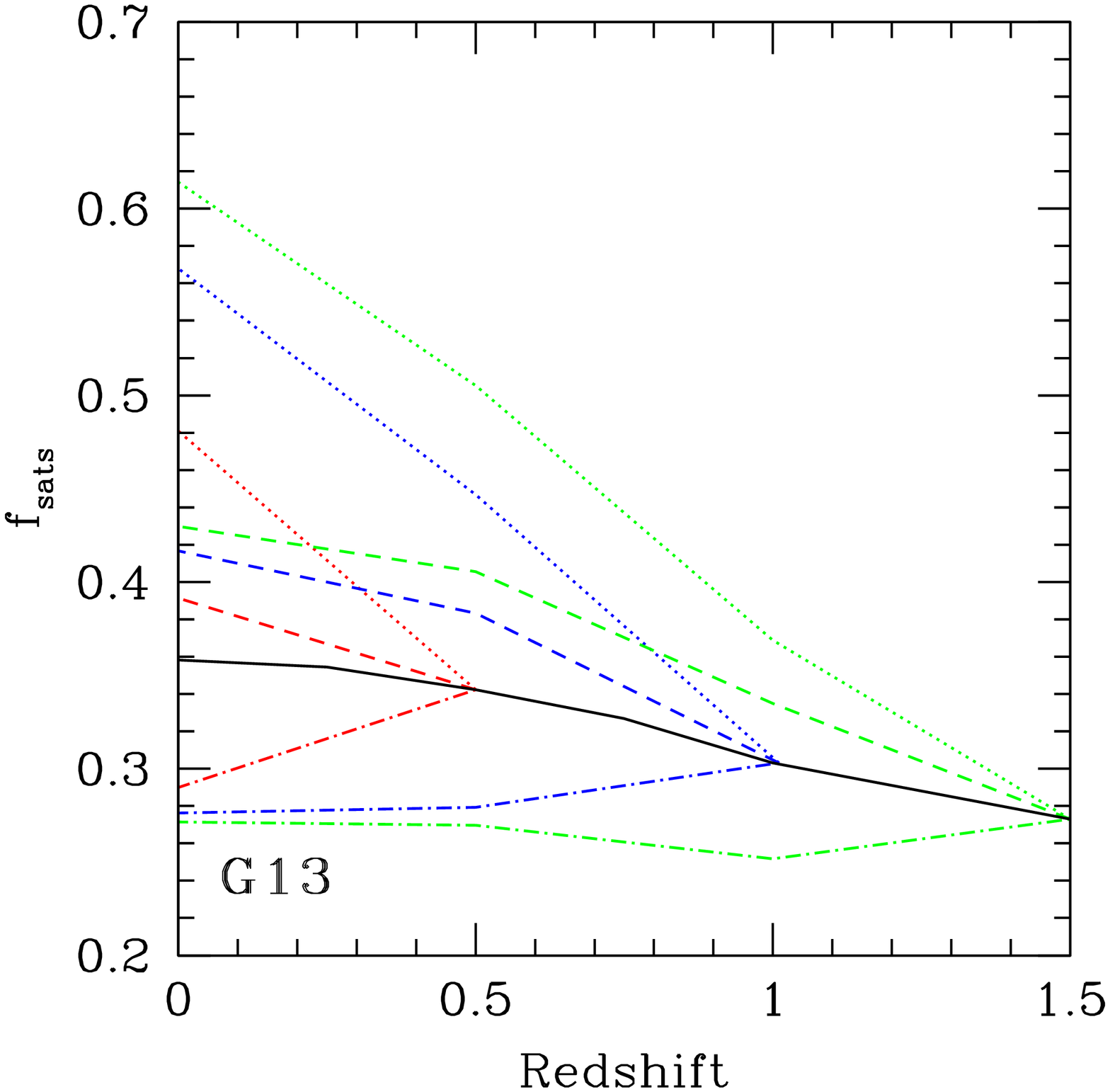}
\caption{
The evolution of the  $M_1/M_{\rm min}$ ratio (top) and the satellite fraction, 
$f_{sat}$, (bottom) in the G13 model for the fixed number density sample of 
$6.56\times 10^{-3} \rm h^{3} Mpc^{-3}$ (solid black lines). 
The other lines show the predictions of the alternative heuristic models 
discussed in the text, for different choices of the selection redshift
(i.e. where they start 
departing from the solid black line): passive evolution (dotted lines), 
tracking evolution (dashed lines) and descendant clustering evolution 
(dashed-dotted lines). The colours indicate the redshift at which these 
models are defined: $z=0.5$ (red), $z=1$ (blue) and $z=1.5$ (green).
}
\label{Fig:PassTrack}
\end{figure}

In Fig.~\ref{Fig:PassTrack} we compare the ratio $M_{1}/M_{\rm min}$ (top panel) 
and the satellite fraction (bottom panel) obtained for the different evolution
models set out above. The heuristic models are defined using different 
selection samples taken from the G13 model, varying the selection redshift 
using a space density of $n = 6.56\times 10^{-3} \,h^{3} \, {\rm Mpc^{-3}}$. 
The black solid line in each panel shows the value of these quantities 
for the fixed number density extracted from the output of the SAMs at each 
redshift. The predictions of the other models are shown for different 
definition redshifts, which correspond to the redshifts at which the other 
line colours and styles branch off the black line.

The predictions of the alternative evolution models shown in 
Fig.~\ref{Fig:PassTrack} for the evolution of the $M_{1}/M_{\rm min}$ ratio
and the satellites fraction are very different from the values measured in 
the SAM output for the fixed number density case.
In particular, in the passive evolution model, satellite galaxies can only
be accreted over time, but not destroyed. This leads to a dramatic increase
of the satellite fraction with time (going toward smaller redshifts) and
a decrease of the $M_{1}/M_{\rm min}$ ratio, in agreement with the conclusions
reached by \cite{Seo:2008}. Thus, observing the opposite trend, as predicted 
by the SAMs for the fixed number density case, can serve as a clear diagnostic
for non-passive evolution of the galaxies.
The tracking model, in which satellites accrete and merge over time
while no new galaxies enter the evolving sample, results in a shallower 
increase of the satellite fraction and a shallower decline of the 
characteristic masses ratio.
The descendants model predictions for the $M_{1}/M_{\rm min}$ ratio are similar 
to the passive evolution one, while the satellite fraction decreases with time. 
These trends are due to the descendants model producing a lower number 
density at z=0 than the higher-redshift starting one, which leads to generally
decreasing with time satellite fractions and $M_{1}/M_{\rm min}$ values.

In all the alternative models, the ratio $M_{1}/M_{\rm min}$ is predicted 
to decline with time following the selection 
redshift, as shown in Fig.~\ref{Fig:PassTrack}, whereas for the fixed
number density samples in the SAM output this ratio increases by $\sim 50\%$ 
by the present day, over the redshift interval plotted. 
This, again, highlights the importance of this diagnostic in deciphering 
among different evolution scenarios.

The predictions of the empirical models for the satellite fraction can 
diverge in either  direction away from the SAM output, to both higher and 
lower values. The reason that the passive evolution model predicts a 
substantially larger number of satellites than is seen in the SAM model 
output (and as a consequence, a lower value of $M_1/M_{\rm min}$) 
is because galaxy mergers are not allowed in this model, which preserves 
the number of galaxies. However, halo mergers do take place with the 
consequence that central galaxies are converted into 
satellite galaxies when their host halo merges with a more massive halo. 
Even with galaxy mergers occurring in the case of the fixed number density 
evolution, it appears that the balance is toward converting centrals to
satellites, such that the satellite fraction increases mildly with time.
This trend is also in accordance with observational estimates (e.g., 
\citealt{Zheng:2007,Coupon:2012,Skibba:2015}).

\begin{figure}
\includegraphics[width=0.50\textwidth]{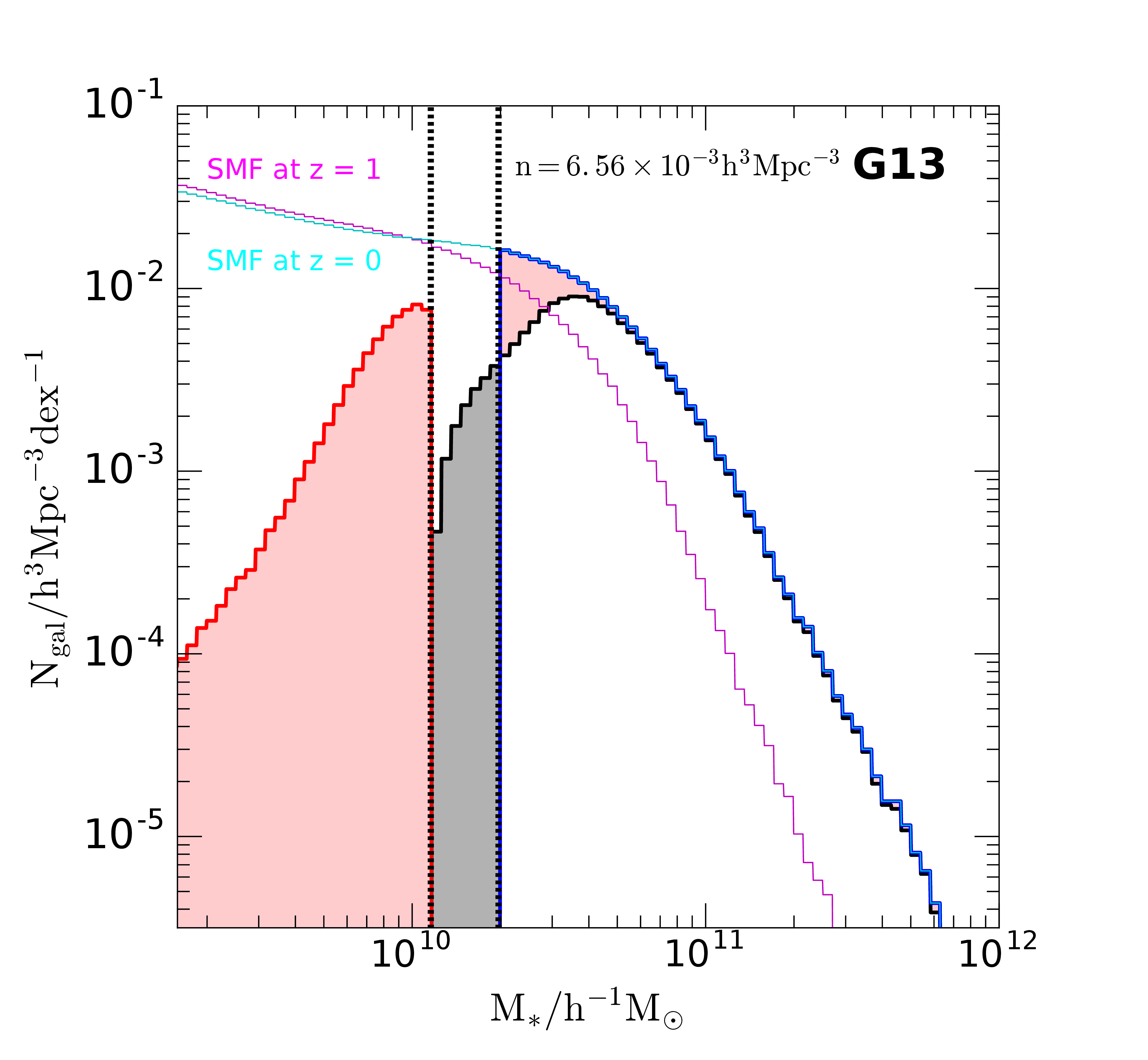}
\includegraphics[width=0.50\textwidth]{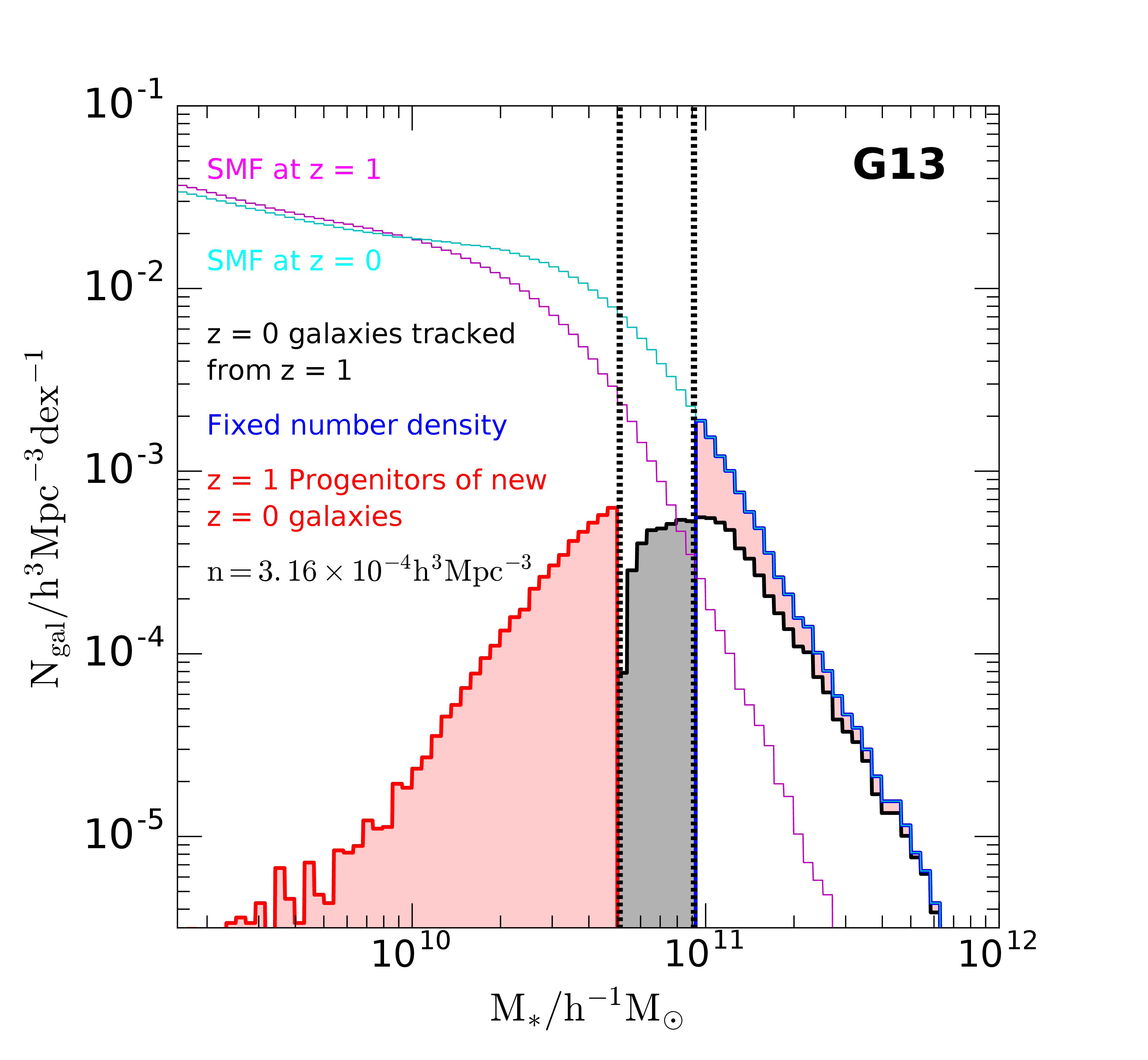}
\caption{The changing membership of stellar-mass selected samples in the 
G13 model, for a number density of $6.56\times 10^{-3} \, h^{-3} \, {\rm Mpc}^{3}$ (top) and $n =  3.16\times 10^{-4}\rm h^{3} Mpc^{-3}$ (bottom). The overall 
stellar mass function is shown for $z=0$ (cyan) and $z=1$ (magenta). The 
black dashed vertical lines show the cut in stellar mass for the number 
density sample at $z=1$ (leftmost line) and $z=0$ (rightmost line). 
The blue solid line histogram represents the stellar mass distribution of
galaxies that make up  the stated fixed number density sample at $z=0$ ( rightmost histogram line). The 
black solid line histogram shows the galaxies at $z=0$ whose progenitors 
consisted of the fixed number density sample at $z=1$ ( middle histogram line). The white region
in common of these two histograms represent the galaxies who remained in 
the sample from $z=1$ to $z=0$, while the grey shaded area represents
the descendants of the galaxies that were in the sample at $z=1$ but are 
no longer members at $z=0$ ( middle shaded area). Also, conversely, many progenitors of the galaxies 
that are in the sample at $z=0$ were not members at $z=1$. These are denoted
by the shaded pink areas ( rightmost and leftmost shaded areas), 
where the red line histogram shows their distribution
at $z=1$ ( leftmost histogram line), and the shaded pink region under the blue histogram shows their
stellar mass at $z=0$. These galaxies compensate for both the grey-region 
galaxies that fell out of the sample due to stunted stellar mass growth and 
for the galaxies that were destroyed due to merging since $z=1$.}
\label{Fig:SMFE}
\end{figure}

Another major difference between the evolution of the fixed number density 
samples and the passive and tracking models  is the significant change in the
identity of the galaxies in the sample in the former case. The change
goes much beyond compensating for the galaxy mergers that occur over time.
A large number of galaxies enter and leave the sample essentially due to galaxy
evolution, namely differences in the star formation rates or growth of stellar
mass, that result in changed ranking of the galaxies by stellar mass
\citep{Leja:2013,Mundy:2015,Torrey:2015}.

To illustrate how the membership of the sample changes with redshift,
we show in Fig.~\ref{Fig:SMFE} the evolution in stellar mass between 
$z=1$ and $z=0$ for two representative fixed number density samples in 
the G13 SAM. The galaxies under the histogram that is shaded grey were 
originally members of the sample defined by number density at $z=1$, but 
are no longer part of a sample defined by the same number density at $z=0$. 
These galaxies are no longer the most massive, but instead have been 
replaced by the galaxies which, at $z=1$, corresponded to those shaded 
pink under the under the red histogram. These galaxies were not massive 
enough to be included in the sample at $z=1$ but grew in mass more quickly 
than some of the galaxies which were in the sample, hence replacing them 
when the sample was redefined in terms of the number density 
of the most massive galaxies when ranked by stellar mass at $z=0$. 
These promoted galaxies are shown in pink shading under the blue histogram. 
(Note when comparing the areas under the curves that this is a log-log plot.) 
The new galaxies that entered the sample since $z=1$ compensate for both
the ones that fell out of the sample due to stunted stellar mass growth
and the ones that got destroyed by mergers.

\begin{table}
\caption{Evolution of the galaxy samples identity from $z=1$ to $z=0$ for
two representative number densities, $n = 6.56\times 10^{-3} \, h^{-3} \, {\rm Mpc}$ and $n = 3.16\times 10^{-4} \, h^{-3} {\rm Mpc}^3$,  in the G13 SAM.
This table shows the percent of galaxies at $z=1$ that exit the sample during 
the redshift interval and those that entered the sample to maintain the 
constant number of galaxy members. The difference in the percent of galaxies 
that ``enter'' and ``exit'' the sample is equal to the percent of galaxies 
that underwent a merger in that time frame. 
The columns are ``total galaxies''  representing the full galaxy sample 
and ``satellite galaxies'', representing the percent (of the the  total 
number of galaxies in the sample at $z=1$) of galaxies that are satellites 
(i.e., the percent that are central galaxies equals ``total galaxies'' 
minus ``satellite galaxies'' numbers).}
\label{tab:100}
\begin{center}
\begin{tabular}{ |c|l|c|c| }
\hline
 $\rm n/h^{-3}Mpc^3$ & Status & Total Galaxies & Satellite Galaxies \\ \hline
 $6.56\times10^{-3}$ & Exit & 8 & 5 \\ 
 $6.56\times10^{-3}$ & Enter & 33 & 10 \\
\hline
$3.16\times10^{-4}$ & Exit & 34 & 13 \\ 
$3.16\times10^{-4}$ & Enter & 55 & 13 \\
\hline
\end{tabular}
\end{center}
\end{table}

These changes in the sample identity are in fact quite significant.
Table~\ref{tab:100} provides the percent of galaxies that exit and enter
these two fixed number density samples between $z=1$ and $z=0$. The difference 
between the galaxies that enter and leave is equal to the number of galaxies 
that merge with other members of the sample. For the $n = 6.56\times 10^{-3} \, h^{-3} \, {\rm Mpc}$  sample, 8\% of the galaxies exit the sample between 
$z=1$
and $z=0$ and about a third of the sample are new galaxies that entered.
The turn in membership is even more prominent for lower number densities (more
massive galaxies), and for the $n = 3.16\times 10^{-4} \, h^{-3} {\rm Mpc}^3$ 
sample about a third of the galaxies exit and more than half of the galaxies
enter the sample over that redshift interval.

These changes in the sample identity also impact the number of central and 
satellite galaxies, though the more significant factor is the 
balance between accretion and destruction of satellites within the sample.
That is, satellite galaxies which merge with their central galaxies tend to
be replaced by central galaxies (whose stellar mass growth rate is typically
larger than satellites that experience quenching). However, the dominant 
effect seems to be the halo mergers turning central galaxies into satellites,
resulting in an overall slight increase of the satellite fraction with time.

\section{Summary \& Conclusions}
\label{Sec:Concl}

The halo occupation distribution (HOD) framework has proven to be a useful
theoretical tool to interpret galaxy clustering measurements and describe
the relation between galaxies and dark matter haloes. Here we
set to study how the halo occupation models evolve with time, an aspect that 
is missing from standard applications, using the outputs of semi-analytic
models (SAMs) that capture the galaxy formation physics.
It is important to recall that the SAMs  predict the galaxy content of dark 
matter haloes along with the properties of these galaxies. The halo occupation
functions are used here as a useful approach to characterise how the haloes
are populated by galaxies in the SAMs.
The halo occupation function has the attraction that it can be readily written in terms 
of the contribution from the main (e.g. most luminous or the galaxy from the 
most massive progenitor halo) or central galaxy, and satellite galaxies, which 
were once central galaxies in their host haloes but have subsequently merged with 
more massive dark matter haloes. Furthermore, the halo occupation function in 
itself is not dependent 
on the radial distribution of galaxies within haloes (though an assumption about 
this is required to predict the correlation function from the HOD). This is 
appealing for our purposes, as different SAMs handle the placement of galaxies 
within haloes in different ways (see the discussion in \citealt{C13} and 
\citealt{Campbell:2015}). 

The SAMs we consider use different implementations of the physical processes 
involved in galaxy formation and set the values of the model parameters in 
different ways, putting emphasis on different observables \citep{Henriques:2015, 
Lacey:2016}. 
We compare the model output at a series of number densities for galaxies ranked 
by their stellar mass. (Note that we also show the stellar mass functions so the 
reader can see how closely these agree with one another.) The HODs look remarkably 
similar until the samples characterised by the lowest number densities. In this 
case, the details of the suppression of gas cooling by heating by accretion onto 
active galactic nuclei become important and introduce differences in the HOD of 
central galaxies. 

The main aim for this study is to characterise the evolution of the HOD 
at a fixed number density, and we explore the evolution of the HOD best-fitting
parameters over the redshift range $0 \le z \le 3$.
As always, it is important to first assess which features of the SAMs are 
robust to the details of the implementation of the physics and the setting 
of the model parameters. Four out of the five parameters in the HOD 
parametrization that we used displayed remarkably similar behaviour. 
As before this similarity was strained when comparing the lowest 
density samples or the parameter which describes the transition from 
zero to one galaxy for the central HOD. 

Three of the HOD parameters are masses (see Fig.~1 for an illustration 
of how the parameters control the shape of the HOD). The evolution of the 
best-fitting values of these masses for samples of fixed number density 
is much weaker than the evolution in the characteristic halo mass (roughly 
speaking the mass at which there is a break from a power-law in the halo 
mass function). We found that the evolution is well described by a single 
parameter describing a power law in redshift and the $z=0$ value of the 
parameter. 

We also compared the evolution predicted in the SAMs to simplified evolution 
models that have been used to model galaxy clustering and evolution. These 
models make different assumptions about the fate of ``galaxies'' identified 
at some redshift. None of these models behave in the same way as the output 
of the SAMs, giving very different predictions for the evolution of the 
HOD parameters and the fraction of satellite galaxies in the sample. 
We find, in particular, that the ratio between the characteristic halo mass
for hosting a satellite galaxy to that of hosting a central galaxy and its
change with redshift can serve as a sensitive diagnostic for different galaxy 
formation and evolution scenarios.

In so far as the models describe the clustering of stellar mass selected 
samples and its evolution, our results can be used to build mock catalogues 
for surveys from $z=0$ to $z=3$.  
Typically, an observational determination of the HOD 
may exist for one redshift, eg the low redshift results for $r$-band selected 
galaxies from \cite{Zehavi:2011}. The problem becomes how to extend these 
best-fitting parameters to other redshifts where there may not be an equivalent 
determination of the HOD parameters. 
For example, one might want to build a mock catalogue for the Euclid redshift 
survey, which will recover emission line galaxies over the redshift range 
$z \approx 0.5 - 2$ from a measurement of the clustering of H-alpha emitters 
at a different redshift (e.g. \citealt{Geach:2012}). We plan to pursue such
efforts in future work.

\section*{Acknowledgements}
This work was made possible by the efforts of Gerard Lemson and
colleagues at the German Astronomical Virtual Observatory in setting
up the Millennium Simulation database in Garching, and John
Helly and Lydia Heck in setting up the mirror at Durham. We thank
Zheng Zheng and Hong Guo for many useful discussions. 
We thank the referee for insightful comments that improved the presentation
of the paper.
SC and IZ acknowledge 
the hospitality of the ICC at Durham and the helpful conversations with many
of its members. 
We acknowledge support from the European Commission’s Framework Programme
7, through the Marie Curie International Research Staff Exchange Scheme 
LACEGAL (PIRSES-GA-2010-269264) and from a STFC/Newton Fund award (ST/M007995/1).
SC further acknowledges support from CONICYT Doctoral Fellowship Programme. 
IZ is supported by NSF grant AST-1612085, as well as by a CWRU ACES+ ADVANCE 
Opportunity Grant and a Faculty Seed Grant.
IZ \& PN acknowledge support from the European Research Council, through the
ERC Starting Grant DEGAS-259586.
IZ, CMB \& PN additionally acknowledge the support of the Science and Technology
Facilities Council (ST/L00075X/1).
CMB acknowledges a research fellowship from the Leverhume Trust. 
NP is supported by ``Centro de Astronomıa y Tecnologıas Afines'' BASAL PFB-06
and by Fondecyt Regular 1150300. 
PN further acknowledges the support of the Royal Society through the award
of a University Research Fellowship. 
The calculations for this paper were performed on the ICC Cosmology
Machine, which is part of the DiRAC-2 Facility jointly funded
by STFC, the Large Facilities Capital Fund of BIS, and Durham
University and on the Geryon computer at the Center for 
Astro-Engineering UC, part of the BASAL PFB-06, which received additional
funding from QUIMAL 130008 and Fondequip AIC-57 for upgrades.
\bibliography{Biblio}

\begin{thebibliography}{98}
\expandafter\ifx\csname natexlab\endcsname\relax\def\natexlab#1{#1}\fi

\bibitem[{{Abbas} {et~al}\mbox{.}(2010){Abbas}, {de la Torre}, {Le F{\`e}vre},
  {Guzzo}, {Marinoni}, {Meneux}, {Pollo}, {Zamorani}, {Bottini}, {Garilli}, {Le
  Brun}, {Maccagni}, {Scaramella}, {Scodeggio}, {Tresse}, {Vettolani},
  {Zanichelli}, {Adami}, {Arnouts}, {Bardelli}, {Bolzonella}, {Cappi},
  {Charlot}, {Ciliegi}, {Contini}, {Foucaud}, {Franzetti}, {Gavignaud},
  {Ilbert}, {Iovino}, {Lamareille}, {McCracken}, {Marano}, {Mazure}, {Merighi},
  {Paltani}, {Pell{\`o}}, {Pozzetti}, {Radovich}, {Vergani}, {Zucca}, {Bondi},
  {Bongiorno}, {Brinchmann}, {Cucciati}, {de Ravel}, {Gregorini},
  {Perez-Montero}, {Mellier}, \& {Merluzzi}}]{Abbas:2010}
{Abbas} U. {et~al.}, 2010, \mnras, 406, 1306

\bibitem[{{Baugh}(2006)}]{Baugh:2006}
{Baugh} C.~M., 2006, Reports on Progress in Physics, 69, 3101

\bibitem[{{Behroozi}, {Conroy} \& {Wechsler}(2010){Behroozi}, {Conroy}, \&
  {Wechsler}}]{Behroozi:2010}
{Behroozi} P.~S., {Conroy} C., {Wechsler} R.~H., 2010, \apj, 717, 379

\bibitem[{{Benson}(2010)}]{Benson:2010}
{Benson} A.~J., 2010, \physrep, 495, 33

\bibitem[{{Benson} {et~al}\mbox{.}(2000){Benson}, {Cole}, {Frenk}, {Baugh}, \&
  {Lacey}}]{Benson:2000}
{Benson} A.~J., {Cole} S., {Frenk} C.~S., {Baugh} C.~M., {Lacey} C.~G., 2000,
  \mnras, 311, 793

\bibitem[{{Berlind} \& {Weinberg}(2002)}]{Berlind:2002}
{Berlind} A.~A., {Weinberg} D.~H., 2002, \apj, 575, 587

\bibitem[{{Berlind} {et~al}\mbox{.}(2003){Berlind}, {Weinberg}, {Benson},
  {Baugh}, {Cole}, {Dav{\'e}}, {Frenk}, {Jenkins}, {Katz}, \&
  {Lacey}}]{Berlind:2003}
{Berlind} A.~A. {et~al.}, 2003, \apj, 593, 1

\bibitem[{{Blake}, {Collister} \& {Lahav}(2008){Blake}, {Collister}, \&
  {Lahav}}]{Blake:2008}
{Blake} C., {Collister} A., {Lahav} O., 2008, \mnras, 385, 1257

\bibitem[{{Bower} {et~al}\mbox{.}(2006){Bower}, {Benson}, {Malbon}, {Helly},
  {Frenk}, {Baugh}, {Cole}, \& {Lacey}}]{Bower:2006}
{Bower} R.~G., {Benson} A.~J., {Malbon} R., {Helly} J.~C., {Frenk} C.~S.,
  {Baugh} C.~M., {Cole} S., {Lacey} C.~G., 2006, \mnras, 370, 645

\bibitem[{{Brown} {et~al}\mbox{.}(2008){Brown}, {Zheng}, {White}, {Dey},
  {Jannuzi}, {Benson}, {Brand}, {Brodwin}, \& {Croton}}]{Brown:2008}
{Brown} M.~J.~I. {et~al.}, 2008, \apj, 682, 937

\bibitem[{{Bullock}, {Wechsler} \& {Somerville}(2002){Bullock}, {Wechsler}, \&
  {Somerville}}]{Bullock:2002}
{Bullock} J.~S., {Wechsler} R.~H., {Somerville} R.~S., 2002, \mnras, 329, 246

\bibitem[{{Campbell} {et~al}\mbox{.}(2015){Campbell}, {Baugh}, {Mitchell},
  {Helly}, {Gonzalez-Perez}, {Lacey}, {Lagos}, {Simha}, \&
  {Farrow}}]{Campbell:2015}
{Campbell} D.~J.~R. {et~al.}, 2015, \mnras, 452, 852

\bibitem[{{Cole} {et~al}\mbox{.}(2000){Cole}, {Lacey}, {Baugh}, \&
  {Frenk}}]{Cole:2000}
{Cole} S., {Lacey} C.~G., {Baugh} C.~M., {Frenk} C.~S., 2000, \mnras, 319, 168

\bibitem[{{Conroy}, {Wechsler} \& {Kravtsov}(2006){Conroy}, {Wechsler}, \&
  {Kravtsov}}]{Conroy:2006}
{Conroy} C., {Wechsler} R.~H., {Kravtsov} A.~V., 2006, \apj, 647, 201

\bibitem[{{Contreras} {et~al}\mbox{.}(2013){Contreras}, {Baugh}, {Norberg}, \&
  {Padilla}}]{C13}
{Contreras} S., {Baugh} C.~M., {Norberg} P., {Padilla} N., 2013, \mnras, 432,
  2717

\bibitem[{{Contreras} {et~al}\mbox{.}(2015){Contreras}, {Baugh}, {Norberg}, \&
  {Padilla}}]{C15}
{Contreras} S., {Baugh} C.~M., {Norberg} P., {Padilla} N., 2015, \mnras, 452,
  1861

\bibitem[{{Cooray}(2006)}]{Cooray:2006}
{Cooray} A., 2006, \mnras, 365, 842

\bibitem[{{Cooray} \& {Sheth}(2002)}]{Cooray:2002}
{Cooray} A., {Sheth} R., 2002, \physrep, 372, 1

\bibitem[{{Coupon} {et~al}\mbox{.}(2015){Coupon}, {Arnouts}, {van Waerbeke},
  {Moutard}, {Ilbert}, {van Uitert}, {Erben}, {Garilli}, {Guzzo}, {Heymans},
  {Hildebrandt}, {Hoekstra}, {Kilbinger}, {Kitching}, {Mellier}, {Miller},
  {Scodeggio}, {Bonnett}, {Branchini}, {Davidzon}, {De Lucia}, {Fritz}, {Fu},
  {Hudelot}, {Hudson}, {Kuijken}, {Leauthaud}, {Le F{\`e}vre}, {McCracken},
  {Moscardini}, {Rowe}, {Schrabback}, {Semboloni}, \& {Velander}}]{Coupon:2015}
{Coupon} J. {et~al.}, 2015, \mnras, 449, 1352

\bibitem[{{Coupon} {et~al}\mbox{.}(2012){Coupon}, {Kilbinger}, {McCracken},
  {Ilbert}, {Arnouts}, {Mellier}, {Abbas}, {de la Torre}, {Goranova},
  {Hudelot}, {Kneib}, \& {Le F{\`e}vre}}]{Coupon:2012}
{Coupon} J. {et~al.}, 2012, \aap, 542, A5

\bibitem[{{Croton} {et~al}\mbox{.}(2006){Croton}, {Springel}, {White}, {De
  Lucia}, {Frenk}, {Gao}, {Jenkins}, {Kauffmann}, {Navarro}, \&
  {Yoshida}}]{Croton:2006}
{Croton} D.~J. {et~al.}, 2006, \mnras, 365, 11

\bibitem[{{Croton} {et~al}\mbox{.}(2016){Croton}, {Stevens}, {Tonini}, {Garel},
  {Bernyk}, {Bibiano}, {Hodkinson}, {Mutch}, {Poole}, \&
  {Shattow}}]{Croton:2016}
{Croton} D.~J. {et~al.}, 2016, \apjs, 222, 22

\bibitem[{{Davis} {et~al}\mbox{.}(1985){Davis}, {Efstathiou}, {Frenk}, \&
  {White}}]{Davis:1985}
{Davis} M., {Efstathiou} G., {Frenk} C.~S., {White} S. D.~M., 1985, \apj, 292,
  371

\bibitem[{{de la Torre} {et~al}\mbox{.}(2013){de la Torre}, {Guzzo}, {Peacock},
  {Branchini}, {Iovino}, {Granett}, {Abbas}, {Adami}, {Arnouts}, {Bel},
  {Bolzonella}, {Bottini}, {Cappi}, {Coupon}, {Cucciati}, {Davidzon}, {De
  Lucia}, {Fritz}, {Franzetti}, {Fumana}, {Garilli}, {Ilbert}, {Krywult}, {Le
  Brun}, {Le F{\`e}vre}, {Maccagni}, {Ma{\l}ek}, {Marulli}, {McCracken},
  {Moscardini}, {Paioro}, {Percival}, {Polletta}, {Pollo}, {Schlagenhaufer},
  {Scodeggio}, {Tasca}, {Tojeiro}, {Vergani}, {Zanichelli}, {Burden}, {Di
  Porto}, {Marchetti}, {Marinoni}, {Mellier}, {Monaco}, {Nichol}, {Phleps},
  {Wolk}, \& {Zamorani}}]{delaTorre:2013}
{de la Torre} S. {et~al.}, 2013, \aap, 557, A54

\bibitem[{{De Lucia} \& {Blaizot}(2007)}]{DeLucia:2007}
{De Lucia} G., {Blaizot} J., 2007, \mnras, 375, 2

\bibitem[{{De Lucia}, {Kauffmann} \& {White}(2004){De Lucia}, {Kauffmann}, \&
  {White}}]{DeLucia:2004}
{De Lucia} G., {Kauffmann} G., {White} S.~D.~M., 2004, \mnras, 349, 1101

\bibitem[{{Durkalec} {et~al}\mbox{.}(2015){Durkalec}, {Le F{\`e}vre}, {Pollo},
  {de la Torre}, {Cassata}, {Garilli}, {Le Brun}, {Lemaux}, {Maccagni},
  {Pentericci}, {Tasca}, {Thomas}, {Vanzella}, {Zamorani}, {Zucca},
  {Amor{\'{\i}}n}, {Bardelli}, {Cassar{\`a}}, {Castellano}, {Cimatti},
  {Cucciati}, {Fontana}, {Giavalisco}, {Grazian}, {Hathi}, {Ilbert}, {Paltani},
  {Ribeiro}, {Schaerer}, {Scodeggio}, {Sommariva}, {Talia}, {Tresse},
  {Vergani}, {Capak}, {Charlot}, {Contini}, {Cuby}, {Dunlop}, {Fotopoulou},
  {Koekemoer}, {L{\'o}pez-Sanjuan}, {Mellier}, {Pforr}, {Salvato}, {Scoville},
  {Taniguchi}, \& {Wang}}]{Durkalec:2015}
{Durkalec} A. {et~al.}, 2015, \aap, 583, A128

\bibitem[{{Eisenstein} {et~al}\mbox{.}(2011){Eisenstein}, {Weinberg}, {Agol},
  {Aihara}, {Allende Prieto}, {Anderson}, {Arns}, {Aubourg}, {Bailey},
  {Balbinot}, \& et~al.}]{Eisenstein:2011}
{Eisenstein} D.~J. {et~al.}, 2011, \aj, 142, 72

\bibitem[{{Font} {et~al}\mbox{.}(2008){Font}, {Bower}, {McCarthy}, {Benson},
  {Frenk}, {Helly}, {Lacey}, {Baugh}, \& {Cole}}]{Font:2008}
{Font} A.~S. {et~al.}, 2008, \mnras, 289, 1619

\bibitem[{{Fontanot} {et~al}\mbox{.}(2009){Fontanot}, {De Lucia}, {Monaco},
  {Somerville}, \& {Santini}}]{Fontanot:2009}
{Fontanot} F., {De Lucia} G., {Monaco} P., {Somerville} R.~S., {Santini} P.,
  2009, \mnras, 397, 1776

\bibitem[{{Fry}(1996)}]{Fry:1996}
{Fry} J.~N., 1996, \apjl, 461, L65

\bibitem[{{Geach} {et~al}\mbox{.}(2012){Geach}, {Sobral}, {Hickox}, {Wake},
  {Smail}, {Best}, {Baugh}, \& {Stott}}]{Geach:2012}
{Geach} J.~E., {Sobral} D., {Hickox} R.~C., {Wake} D.~A., {Smail} I., {Best}
  P.~N., {Baugh} C.~M., {Stott} J.~P., 2012, \mnras, 426, 679

\bibitem[{{Gonzalez-Perez} {et~al}\mbox{.}(2014){Gonzalez-Perez}, {Lacey},
  {Baugh}, {Lagos}, {Helly}, {Campbell}, \& {Mitchell}}]{Gonzalez:2014}
{Gonzalez-Perez} V., {Lacey} C.~G., {Baugh} C.~M., {Lagos} C.~D.~P., {Helly}
  J., {Campbell} D.~J.~R., {Mitchell} P.~D., 2014, \mnras, 439, 264

\bibitem[{{Guo} {et~al}\mbox{.}(2015{\natexlab{a}}){Guo}, {Zheng}, {Zehavi},
  {Behroozi}, {Chuang}, {Comparat}, {Favole}, {Gottloeber}, {Klypin}, {Prada},
  {Weinberg}, \& {Yepes}}]{Guo:2015}
{Guo} H. {et~al.}, 2015{\natexlab{a}}, \mnras, 453, 4368

\bibitem[{{Guo} {et~al}\mbox{.}(2015{\natexlab{b}}){Guo}, {Zheng}, {Zehavi},
  {Dawson}, {Skibba}, {Tinker}, {Weinberg}, {White}, \&
  {Schneider}}]{Guo:2015a}
{Guo} H. {et~al.}, 2015{\natexlab{b}}, \mnras, 446, 578

\bibitem[{{Guo} {et~al}\mbox{.}(2014){Guo}, {Zheng}, {Zehavi}, {Xu},
  {Eisenstein}, {Weinberg}, {Bahcall}, {Berlind}, {Comparat}, {McBride},
  {Ross}, {Schneider}, {Skibba}, {Swanson}, {Tinker}, {Tojeiro}, \&
  {Wake}}]{Guo:2014b}
{Guo} H. {et~al.}, 2014, \mnras, 441, 2398

\bibitem[{{Guo} \& {White}(2014)}]{Guo:2014}
{Guo} Q., {White} S., 2014, \mnras, 437, 3228

\bibitem[{{Guo} {et~al}\mbox{.}(2013){Guo}, {White}, {Angulo}, {Henriques},
  {Lemson}, {Boylan-Kolchin}, {Thomas}, \& {Short}}]{Guo:2013a}
{Guo} Q., {White} S., {Angulo} R.~E., {Henriques} B., {Lemson} G.,
  {Boylan-Kolchin} M., {Thomas} P., {Short} C., 2013, \mnras, 428, 1351

\bibitem[{{Guo} {et~al}\mbox{.}(2011){Guo}, {White}, {Boylan-Kolchin}, {De
  Lucia}, {Kauffmann}, {Lemson}, {Li}, {Springel}, \& {Weinmann}}]{Guo:2011}
{Guo} Q. {et~al.}, 2011, \mnras, 413, 101

\bibitem[{{Hamana} {et~al}\mbox{.}(2006){Hamana}, {Yamada}, {Ouchi}, {Iwata},
  \& {Kodama}}]{Hamana:2006}
{Hamana} T., {Yamada} T., {Ouchi} M., {Iwata} I., {Kodama} T., 2006, \mnras,
  369, 1929

\bibitem[{{Hearin} {et~al}\mbox{.}(2016){Hearin}, {Zentner}, {van den Bosch},
  {Campbell}, \& {Tollerud}}]{Hearin:2016}
{Hearin} A.~P., {Zentner} A.~R., {van den Bosch} F.~C., {Campbell} D.,
  {Tollerud} E., 2016, \mnras

\bibitem[{{Henriques} {et~al}\mbox{.}(2015){Henriques}, {White}, {Thomas},
  {Angulo}, {Guo}, {Lemson}, {Springel}, \& {Overzier}}]{Henriques:2015}
{Henriques} B.~M.~B., {White} S.~D.~M., {Thomas} P.~A., {Angulo} R., {Guo} Q.,
  {Lemson} G., {Springel} V., {Overzier} R., 2015, \mnras, 451, 2663

\bibitem[{{Henriques} {et~al}\mbox{.}(2013){Henriques}, {White}, {Thomas},
  {Angulo}, {Guo}, {Lemson}, \& {Springel}}]{Henriques:2013}
{Henriques} B.~M.~B., {White} S.~D.~M., {Thomas} P.~A., {Angulo} R.~E., {Guo}
  Q., {Lemson} G., {Springel} V., 2013, \mnras, 431, 3373

\bibitem[{{Jiang} {et~al}\mbox{.}(2014){Jiang}, {Helly}, {Cole}, \&
  {Frenk}}]{Jiang:2014}
{Jiang} L., {Helly} J.~C., {Cole} S., {Frenk} C.~S., 2014, \mnras, 440, 2115

\bibitem[{{Jing} \& {B{\"o}rner}(1998)}]{Jing:1998b}
{Jing} Y.~P., {B{\"o}rner} G., 1998, \apj, 503, 37

\bibitem[{{Jing}, {B{\"o}rner} \& {Suto}(2002){Jing}, {B{\"o}rner}, \&
  {Suto}}]{Jing:2002}
{Jing} Y.~P., {B{\"o}rner} G., {Suto} Y., 2002, \apj, 564, 15

\bibitem[{{Jing}, {Mo} \& {B{\"o}rner}(1998){Jing}, {Mo}, \&
  {B{\"o}rner}}]{Jing:1998a}
{Jing} Y.~P., {Mo} H.~J., {B{\"o}rner} G., 1998, \apj, 494, 1

\bibitem[{{Kim} {et~al}\mbox{.}(2009){Kim}, {Baugh}, {Cole}, {Frenk}, \&
  {Benson}}]{Kim:2009}
{Kim} H.-S., {Baugh} C.~M., {Cole} S., {Frenk} C.~S., {Benson} A.~J., 2009,
  \mnras, 400, 1527

\bibitem[{{Kim} {et~al}\mbox{.}(2015){Kim}, {Im}, {Lee}, {Edge}, {Wake},
  {Merson}, \& {Jeon}}]{Kim:2015}
{Kim} J.-W., {Im} M., {Lee} S.-K., {Edge} A.~C., {Wake} D.~A., {Merson} A.~I.,
  {Jeon} Y., 2015, \apj, 806, 189

\bibitem[{{Krause} {et~al}\mbox{.}(2013){Krause}, {Hirata}, {Martin}, {Neill},
  \& {Wyder}}]{Krause:2013}
{Krause} E., {Hirata} C.~M., {Martin} C., {Neill} J.~D., {Wyder} T.~K., 2013,
  \mnras, 428, 2548

\bibitem[{{Kravtsov} {et~al}\mbox{.}(2004){Kravtsov}, {Berlind}, {Wechsler},
  {Klypin}, {Gottl{\"o}ber}, {Allgood}, \& {Primack}}]{Kravtsov:2004}
{Kravtsov} A.~V., {Berlind} A.~A., {Wechsler} R.~H., {Klypin} A.~A.,
  {Gottl{\"o}ber} S., {Allgood} B., {Primack} J.~R., 2004, \apj, 609, 35

\bibitem[{{Lacey} {et~al}\mbox{.}(2016){Lacey}, {Baugh}, {Frenk}, {Benson},
  {Bower}, {Cole}, {Gonzalez-Perez}, {Helly}, {Lagos}, \&
  {Mitchell}}]{Lacey:2016}
{Lacey} C.~G. {et~al.}, 2016, \mnras, 462, 3854

\bibitem[{{Lagos} {et~al}\mbox{.}(2011){Lagos}, {Baugh}, {Lacey}, {Benson},
  {Kim}, \& {Power}}]{Lagos:2011b}
{Lagos} C. D.~P., {Baugh} C.~M., {Lacey} C.~G., {Benson} A.~J., {Kim} H.-S.,
  {Power} C., 2011, \mnras, 418, 1649

\bibitem[{{Lagos} {et~al}\mbox{.}(2012){Lagos}, {Bayet}, {Baugh}, {Lacey},
  {Bell}, {Fanidakis}, \& {Geach}}]{Lagos:2012}
{Lagos} C. d.~P., {Bayet} E., {Baugh} C.~M., {Lacey} C.~G., {Bell} T.~A.,
  {Fanidakis} N., {Geach} J.~E., 2012, \mnras, 426, 2142

\bibitem[{{Lee} {et~al}\mbox{.}(2009){Lee}, {Giavalisco}, {Conroy}, {Wechsler},
  {Ferguson}, {Somerville}, {Dickinson}, \& {Urry}}]{Lee:2009}
{Lee} K.-S., {Giavalisco} M., {Conroy} C., {Wechsler} R.~H., {Ferguson} H.~C.,
  {Somerville} R.~S., {Dickinson} M.~E., {Urry} C.~M., 2009, \apj, 695, 368

\bibitem[{{Lee} {et~al}\mbox{.}(2006){Lee}, {Giavalisco}, {Gnedin},
  {Somerville}, {Ferguson}, {Dickinson}, \& {Ouchi}}]{Lee:2006}
{Lee} K.-S., {Giavalisco} M., {Gnedin} O.~Y., {Somerville} R.~S., {Ferguson}
  H.~C., {Dickinson} M., {Ouchi} M., 2006, \apj, 642, 63

\bibitem[{{Leja} {et~al}\mbox{.}(2013){Leja}, {van Dokkum}, {Momcheva},
  {Brammer}, {Skelton}, {Whitaker}, {Andrews}, {Franx}, {Kriek}, {van der Wel},
  {Bezanson}, {Conroy}, {F{\"o}rster Schreiber}, {Nelson}, \&
  {Patel}}]{Leja:2013}
{Leja} J. {et~al.}, 2013, \apjl, 778, L24

\bibitem[{{Magliocchetti} \& {Porciani}(2003)}]{Magliocchetti:2003}
{Magliocchetti} M., {Porciani} C., 2003, \mnras, 346, 186

\bibitem[{{Manera} {et~al}\mbox{.}(2015){Manera}, {Samushia}, {Tojeiro},
  {Howlett}, {Ross}, {Percival}, {Gil-Mar{\'{\i}}n}, {Brownstein}, {Burden}, \&
  {Montesano}}]{Manera:2015}
{Manera} M. {et~al.}, 2015, \mnras, 447, 437

\bibitem[{{McCracken} {et~al}\mbox{.}(2015){McCracken}, {Wolk}, {Colombi},
  {Kilbinger}, {Ilbert}, {Peirani}, {Coupon}, {Dunlop}, {Milvang-Jensen},
  {Caputi}, {Aussel}, {B{\'e}thermin}, \& {Le F{\`e}vre}}]{McCracken:2015}
{McCracken} H.~J. {et~al.}, 2015, \mnras, 449, 901

\bibitem[{{Merson} {et~al}\mbox{.}(2013){Merson}, {Baugh}, {Helly},
  {Gonzalez-Perez}, {Cole}, {Bielby}, {Norberg}, {Frenk}, {Benson}, {Bower},
  {Lacey}, \& {Lagos}}]{Merson:2013}
{Merson} A.~I. {et~al.}, 2013, \mnras, 429, 556

\bibitem[{{Moustakas} \& {Somerville}(2002)}]{Moustakas:2002}
{Moustakas} L.~A., {Somerville} R.~S., 2002, \apj, 577, 1

\bibitem[{{Mundy}, {Conselice} \& {Ownsworth}(2015){Mundy}, {Conselice}, \&
  {Ownsworth}}]{Mundy:2015}
{Mundy} C.~J., {Conselice} C.~J., {Ownsworth} J.~R., 2015, \mnras, 450, 3696

\bibitem[{{Padilla} {et~al}\mbox{.}(2010){Padilla}, {Christlein}, {Gawiser},
  {Gonz{\'a}lez}, {Guaita}, \& {Infante}}]{Padilla:2010}
{Padilla} N.~D., {Christlein} D., {Gawiser} E., {Gonz{\'a}lez} R.~E., {Guaita}
  L., {Infante} L., 2010, \mnras, 409, 184

\bibitem[{{Padilla} {et~al}\mbox{.}(2014){Padilla}, {Salazar-Albornoz},
  {Contreras}, {Cora}, \& {Ruiz}}]{Padilla:2014}
{Padilla} N.~D., {Salazar-Albornoz} S., {Contreras} S., {Cora} S.~A., {Ruiz}
  A.~N., 2014, \mnras, 443, 2801

\bibitem[{{Parejko} {et~al}\mbox{.}(2013){Parejko}, {Sunayama}, {Padmanabhan},
  {Wake}, {Berlind}, {Bizyaev}, {Blanton}, {Bolton}, {van den Bosch},
  {Brinkmann}, {Brownstein}, {da Costa}, {Eisenstein}, {Guo}, {Kazin}, {Maia},
  {Malanushenko}, {Maraston}, {McBride}, {Nichol}, {Oravetz}, {Pan},
  {Percival}, {Prada}, {Ross}, {Ross}, {Schlegel}, {Schneider}, {Simmons},
  {Skibba}, {Tinker}, {Tojeiro}, {Weaver}, {Wetzel}, {White}, {Weinberg},
  {Thomas}, {Zehavi}, \& {Zheng}}]{Parejko:2013}
{Parejko} J.~K. {et~al.}, 2013, \mnras, 429, 98

\bibitem[{{Peacock} \& {Smith}(2000)}]{Peacock:2000}
{Peacock} J.~A., {Smith} R.~E., 2000, \mnras, 318, 1144

\bibitem[{{Phleps} {et~al}\mbox{.}(2006){Phleps}, {Peacock}, {Meisenheimer}, \&
  {Wolf}}]{Phleps:2006}
{Phleps} S., {Peacock} J.~A., {Meisenheimer} K., {Wolf} C., 2006, \aap, 457,
  145

\bibitem[{{Quadri} {et~al}\mbox{.}(2008){Quadri}, {Williams}, {Lee}, {Franx},
  {van Dokkum}, \& {Brammer}}]{Quadri:2008}
{Quadri} R.~F., {Williams} R.~J., {Lee} K.-S., {Franx} M., {van Dokkum} P.,
  {Brammer} G.~B., 2008, \apjl, 685, L1

\bibitem[{{Rodr{\'{\i}}guez-Puebla}
  {et~al}\mbox{.}(2016){Rodr{\'{\i}}guez-Puebla}, {Behroozi}, {Primack},
  {Klypin}, {Lee}, \& {Hellinger}}]{Rod-Puebla:2016}
{Rodr{\'{\i}}guez-Puebla} A., {Behroozi} P., {Primack} J., {Klypin} A., {Lee}
  C., {Hellinger} D., 2016, \mnras, 462, 893

\bibitem[{{Ross}, {Percival} \& {Brunner}(2010){Ross}, {Percival}, \&
  {Brunner}}]{Ross:2010}
{Ross} A.~J., {Percival} W.~J., {Brunner} R.~J., 2010, \mnras, 407, 420

\bibitem[{{Scoccimarro} {et~al}\mbox{.}(2001){Scoccimarro}, {Feldman}, {Fry},
  \& {Frieman}}]{Scoccimarro:2001}
{Scoccimarro} R., {Feldman} H.~A., {Fry} J.~N., {Frieman} J.~A., 2001, \apj,
  546, 652

\bibitem[{{Seljak}(2000)}]{Seljak:2000}
{Seljak} U., 2000, \mnras, 318, 203

\bibitem[{{Seo}, {Eisenstein} \& {Zehavi}(2008){Seo}, {Eisenstein}, \&
  {Zehavi}}]{Seo:2008}
{Seo} H.-J., {Eisenstein} D.~J., {Zehavi} I., 2008, \apj, 681, 998

\bibitem[{{Simon} {et~al}\mbox{.}(2009){Simon}, {Hetterscheidt}, {Wolf},
  {Meisenheimer}, {Hildebrandt}, {Schneider}, {Schirmer}, \&
  {Erben}}]{Simon:2009}
{Simon} P., {Hetterscheidt} M., {Wolf} C., {Meisenheimer} K., {Hildebrandt} H.,
  {Schneider} P., {Schirmer} M., {Erben} T., 2009, \mnras, 398, 807

\bibitem[{{Skibba} {et~al}\mbox{.}(2015){Skibba}, {Coil}, {Mendez}, {Blanton},
  {Bray}, {Cool}, {Eisenstein}, {Guo}, {Miyaji}, {Moustakas}, \&
  {Zhu}}]{Skibba:2015}
{Skibba} R.~A. {et~al.}, 2015, \apj, 807, 152

\bibitem[{{Skibba} {et~al}\mbox{.}(2014){Skibba}, {Smith}, {Coil}, {Moustakas},
  {Aird}, {Blanton}, {Bray}, {Cool}, {Eisenstein}, {Mendez}, {Wong}, \&
  {Zhu}}]{Skibba:2014}
{Skibba} R.~A. {et~al.}, 2014, \apj, 784, 128

\bibitem[{{Somerville} \& {Dav{\'e}}(2015)}]{Somerville:2015}
{Somerville} R.~S., {Dav{\'e}} R., 2015, \araa, 53, 51

\bibitem[{{Springel} {et~al}\mbox{.}(2005){Springel}, {White}, {Jenkins},
  {Frenk}, {Yoshida}, {Gao}, {Navarro}, {Thacker}, {Croton}, {Helly},
  {Peacock}, \& {Cole}}]{Springel:2005}
{Springel} V. {et~al.}, 2005, \nat, 435, 629

\bibitem[{{Springel} {et~al}\mbox{.}(2001){Springel}, {White}, {Tormen}, \&
  {Kauffmann}}]{Springel:2001}
{Springel} V., {White} S. D.~M., {Tormen} G., {Kauffmann} G., 2001, \mnras,
  328, 726

\bibitem[{{Torrey} {et~al}\mbox{.}(2015){Torrey}, {Wellons}, {Machado},
  {Griffen}, {Nelson}, {Rodriguez-Gomez}, {McKinnon}, {Pillepich}, {Ma},
  {Vogelsberger}, {Springel}, \& {Hernquist}}]{Torrey:2015}
{Torrey} P. {et~al.}, 2015, \mnras, 454, 2770

\bibitem[{{van den Bosch}, {Yang} \& {Mo}(2003){van den Bosch}, {Yang}, \&
  {Mo}}]{vandenBosch:2003}
{van den Bosch} F.~C., {Yang} X., {Mo} H.~J., 2003, \mnras, 340, 771

\bibitem[{{Wake} {et~al}\mbox{.}(2008){Wake}, {Sheth}, {Nichol}, {Baugh},
  {Bland-Hawthorn}, {Colless}, {Couch}, {Croom}, {de Propris}, {Drinkwater},
  {Edge}, {Loveday}, {Lam}, {Pimbblet}, {Roseboom}, {Ross}, {Schneider},
  {Shanks}, \& {Sharp}}]{Wake:2008}
{Wake} D.~A. {et~al.}, 2008, \mnras, 387, 1045

\bibitem[{{Wake} {et~al}\mbox{.}(2011){Wake}, {Whitaker}, {Labb{\'e}}, {van
  Dokkum}, {Franx}, {Quadri}, {Brammer}, {Kriek}, {Lundgren}, {Marchesini}, \&
  {Muzzin}}]{Wake:2011}
{Wake} D.~A. {et~al.}, 2011, \apj, 728, 46

\bibitem[{{Watson}, {Berlind} \& {Zentner}(2011){Watson}, {Berlind}, \&
  {Zentner}}]{Watson:2011}
{Watson} D.~F., {Berlind} A.~A., {Zentner} A.~R., 2011, \apj, 738, 22

\bibitem[{{White} {et~al}\mbox{.}(2011){White}, {Blanton}, {Bolton},
  {Schlegel}, {Tinker}, {Berlind}, {da Costa}, {Kazin}, {Lin}, {Maia},
  {McBride}, {Padmanabhan}, {Parejko}, {Percival}, {Prada}, {Ramos}, {Sheldon},
  {de Simoni}, {Skibba}, {Thomas}, {Wake}, {Zehavi}, {Zheng}, {Nichol},
  {Schneider}, {Strauss}, {Weaver}, \& {Weinberg}}]{White:2011}
{White} M. {et~al.}, 2011, \apj, 728, 126

\bibitem[{{White} {et~al}\mbox{.}(2007){White}, {Zheng}, {Brown}, {Dey}, \&
  {Jannuzi}}]{White:2007}
{White} M., {Zheng} Z., {Brown} M.~J.~I., {Dey} A., {Jannuzi} B.~T., 2007,
  \apjl, 655, L69

\bibitem[{{Yan}, {Madgwick} \& {White}(2003){Yan}, {Madgwick}, \&
  {White}}]{Yan:2003}
{Yan} R., {Madgwick} D.~S., {White} M., 2003, \apj, 598, 848

\bibitem[{{Yang} {et~al}\mbox{.}(2005){Yang}, {Mo}, {Jing}, \& {van den
  Bosch}}]{Yang:2005}
{Yang} X., {Mo} H.~J., {Jing} Y.~P., {van den Bosch} F.~C., 2005, \mnras, 358,
  217

\bibitem[{{Yang}, {Mo} \& {van den Bosch}(2003){Yang}, {Mo}, \& {van den
  Bosch}}]{Yang:2003}
{Yang} X., {Mo} H.~J., {van den Bosch} F.~C., 2003, \mnras, 339, 1057

\bibitem[{{Zehavi} {et~al}\mbox{.}(2011){Zehavi}, {Zheng}, {Weinberg},
  {Blanton}, {Bahcall}, {Berlind}, {Brinkmann}, {Frieman}, {Gunn}, {Lupton},
  {Nichol}, {Percival}, {Schneider}, {Skibba}, {Strauss}, {Tegmark}, \&
  {York}}]{Zehavi:2011}
{Zehavi} I. {et~al.}, 2011, \apj, 736, 59

\bibitem[{{Zehavi} {et~al}\mbox{.}(2005){Zehavi}, {Zheng}, {Weinberg},
  {Frieman}, {Berlind}, {Blanton}, {Scoccimarro}, {Sheth}, {Strauss}, {Kayo},
  {Suto}, {Fukugita}, {Nakamura}, {Bahcall}, {Brinkmann}, {Gunn}, {Hennessy},
  {Ivezi{\'c}}, {Knapp}, {Loveday}, {Meiksin}, {Schlegel}, {Schneider},
  {Szapudi}, {Tegmark}, {Vogeley}, {York}, \& {SDSS
  Collaboration}}]{Zehavi:2005}
{Zehavi} I. {et~al.}, 2005, \apj, 630, 1

\bibitem[{{Zentner} {et~al}\mbox{.}(2005){Zentner}, {Berlind}, {Bullock},
  {Kravtsov}, \& {Wechsler}}]{Zentner:2005}
{Zentner} A.~R., {Berlind} A.~A., {Bullock} J.~S., {Kravtsov} A.~V., {Wechsler}
  R.~H., 2005, \apj, 624, 505

\bibitem[{{Zheng}(2004)}]{Zheng:2004}
{Zheng} Z., 2004, \apj, 610, 61

\bibitem[{{Zheng} {et~al}\mbox{.}(2005){Zheng}, {Berlind}, {Weinberg},
  {Benson}, {Baugh}, {Cole}, {Dav{\'e}}, {Frenk}, {Katz}, \&
  {Lacey}}]{Zheng:2005}
{Zheng} Z. {et~al.}, 2005, \apj, 633, 791

\bibitem[{{Zheng}, {Coil} \& {Zehavi}(2007){Zheng}, {Coil}, \&
  {Zehavi}}]{Zheng:2007}
{Zheng} Z., {Coil} A.~L., {Zehavi} I., 2007, \apj, 667, 760

\bibitem[{{Zheng} \& {Guo}(2016)}]{Zheng:2016}
{Zheng} Z., {Guo} H., 2016, \mnras, 458, 4015

\bibitem[{{Zheng} {et~al}\mbox{.}(2009){Zheng}, {Zehavi}, {Eisenstein},
  {Weinberg}, \& {Jing}}]{Zheng:2009}
{Zheng} Z., {Zehavi} I., {Eisenstein} D.~J., {Weinberg} D.~H., {Jing} Y.~P.,
  2009, \apj, 707, 554

\end{thebibliography}

\label{lastpage}
\end{document}